\documentclass[conference]{IEEEtran}

\usepackage{todonotes}
\usepackage[ruled,linesnumbered]{algorithm2e}
\usepackage{amsmath}%
\usepackage{amssymb}
\usepackage{xspace}
\usepackage{ltl}
\usepackage{tikz}
\usetikzlibrary{positioning, arrows.meta}
\usepackage{caption}
\usepackage{subcaption}
\usepackage{tcolorbox}
\usepackage{hyperref}
\usepackage{csvsimple}
\usepackage{pgfplots}
\usepackage{graphicx}
\pgfplotsset{compat=newest}
\usepackage[]{algorithm2e}
\usepackage{ stmaryrd }
\usepackage{array}
\usepackage{wrapfig}
\usepackage{multirow}
\usepackage[switch]{lineno}
\usepackage{xcolor}
\usepackage{orcidlink}

\hypersetup{%
	colorlinks=true,           
	allcolors=blue!70!black,   
	pdfstartview=Fit,          
	breaklinks=true,
}
\newcommand{\exvars}{\mathcal{U}}
\newcommand{\envars}{\mathcal{V}}
\newcommand{\range}{\mathcal{R}}

\newcommand{\qed}{$\blacksquare$}

\newcommand{\sig}{\mathcal{S}}
\newcommand{\streq}{\mathcal{F}}
\newcommand{\MOD}{\mathcal{M}}
\newcommand{\MU}{(\MOD, \vec{u})}
\newcommand{\intrv}{\mathcal{I}}

\newcommand{\cause}{\mathsf{c}}
\newcommand{\effect}{\mathsf{e}}

\newcommand{\trans}{\mathcal{T}}
\newcommand{\States}{\Sigma}
\newcommand{\Trans}{\Delta}
\newcommand{\Init}{\sigma^0}
\newcommand{\Labels}{\Lambda}
\newcommand{\Traces}{\mathsf{Tr}}

\newcommand{\G}{\LTLsquare}
\newcommand{\F}{\LTLdiamond}
\newcommand{\U}{\mathbin{\mathcal{U}}}

\newcommand{\undr}[1]{\check{#1}}
\newcommand{\ovr}[1]{\hat{#1}}

\newcommand{\fail}{\mathsf{fail}}

\newcommand{\HP}{\textsf{\small HP}\xspace}

\newcommand{\pos}{\mathit{pos}}
\newcommand{\vel}{\mathit{vel}}
\newcommand{\action}{\mathit{action}}

\newcommand{\ENDO}{endogenous\xspace}
\newcommand{\EXO}{exogenous\xspace}
\newcommand{\AP}{\textsf{AP}}

\renewcommand{\notin}{\not\in}
\newtheorem{definition}{Definition}

\newcommand{\FunOvrV}{\ovr{h}}

\newcommand{\hintO}{\ovr{w}}

\newcommand{\Refine}{\mathsf{Refine}}
\newcommand{\FunUndV}{\undr{h}}

\newcommand{\Rst}{\mathsf{Rst}}
\newcommand{\rst}{|}

\newtheorem{example}{Example}
\newtheorem{assumption}{Assumption}
\newtheorem{theorem}{Theorem}

\newcommand{\wind}{\mathit{wind}}
\newcommand{\gravity}{\mathit{gravity}}
\newcommand{\torbu}{\mathit{turbulence}}

\newcommand{\speed}{\mathit{airspeed}}
\newcommand{\alt}{\mathit{altitude}}
\newcommand{\pit}{\mathit{pitch}}
\newcommand{\pow}{\mathit{power \textendash lag}}
\newcommand{\aoa}{\mathit{\alpha}}
\newcommand{\yaw}{\mathit{yaw}}
\newcommand{\roll}{\mathit{roll}}
\newcommand{\throt}{\delta_t}
\newcommand{\elev}{\delta_e}
\newcommand{\ail}{\delta_a}
\newcommand{\rud}{\delta_r}

\definecolor{gray}{rgb}{0.5,0.5,0.5}
\definecolor{darkgreen}{rgb}{0,0.6,0}
\definecolor{darkblue}{rgb}{0.2, 0.2, 0.6}
\definecolor{niceblue}{rgb}{0.16, 0.32, 0.75}
\definecolor{niceblack}{rgb}{0.0, 0.18, 0.39}
\definecolor{prettyred}{rgb}{1.0, 0.13, 0.32}
\definecolor{prettypink}{rgb}{0.98, 0.38, 0.5}
\definecolor{prettyblue}{rgb}{0.0, 0.30, 1.0}
\definecolor{prettygreen}{rgb}{0.0, 0.5, 0.0}
\definecolor{prettyorange}{rgb}{1.0, 0.33, 0.0}
\definecolor{prettypurple}{rgb}{0.6, 0.4, 0.8}
\definecolor{prettyyellow}{rgb}{0.99, 0.93, 0.0}

\newcommand{\revision}[1]{\textcolor{black}{#1}}

\begin{document} 


\title{Efficient Discovery of Actual Causality using Abstraction-Refinement} 

\title{Efficient Discovery of Actual Causality using Abstraction-Refinement
\thanks{This work is sponsored by the United States NSF Award CCF 2320050.}} 

\author{\IEEEauthorblockN{Arshia Rafieioskouei \orcidlink{0009-0002-8844-4441}}
	\IEEEauthorblockA{
		Michigan State University\\
		\url{rafieios@msu.edu}}
	\and
\IEEEauthorblockN{Borzoo Bonakdarpour \orcidlink{0000-0003-1800-5419}}
\IEEEauthorblockA{
	Michigan State University\\
	\url{borzoo@msu.edu }}
	}


\maketitle              

\begin{abstract}
\emph{Causality} is the relationship where one event contributes to the production of another, with the cause being partly responsible for the effect and the effect partly dependent on the cause.
In this paper, we propose a novel and effective method to formally reason about the causal effect of 
events in engineered systems, with \revision{application for finding the root-cause of safety 
violations in embedded and cyber-physical systems}.
We are motivated by the notion of {\em actual causality} by Halpern and Pearl, which focuses on the 
causal effect of particular events rather than type-level causality, which attempts to make general 
statements about scientific and natural phenomena.
Our first contribution is formulating discovery of actual causality in computing systems 
modeled by transition systems as an SMT solving problem.

Since datasets for causality analysis tend to be large, in order to tackle the scalability problem of 
automated formal reasoning, our second contribution is a novel technique based on {\em 
abstraction-refinement} that allows identifying for actual causes within smaller abstract causal models.
We demonstrate the effectiveness of our approach (by several orders of magnitude) using 
\revision{three} case studies to find the actual cause of violations of safety in (1) a neural network 
controller for a Mountain Car, (2) a controller for a Lunar Lander obtained by reinforcement 
learning, and \revision{(3) an MPC controller for an F-16 autopilot simulator.}

 \end{abstract}

\section{Introduction}

In a {\em causal system}, the output of the system is influenced only by the present and past inputs.
In other words, in a causal system, the present and future outputs depend solely on past and 
present inputs, not on future inputs.
Causality addresses the logical dependencies between events and reflects the essence of event and action flows in systems.
Engineers generally build causal systems, that is, structures, systems, and processes that 
seek to tie effects to their causes.
This also includes approaches to explain the root-cause of failures that violate safety standards, 
especially in safety-critical systems.

In this context, embedded and cyber-physical systems (CPS) are no exceptions.
In fact, root-cause analysis has been of interest to both academic and industrial circles for decades, aiming not just to find safety violations but also to precisely {\em explain} why they happened.
That is, one needs to mathematically prove that in the absence of this reason, safety would not have 
been violated.
This means proving mathematically that safety would not have been violated in the absence of the identified cause.
Formalizing and reasoning about causal explanations is much harder than just finding ``bugs'' and 
often aims to identify the earliest flawed decisions by controllers that lead to violations of safety 
requirements.
Put it another way, causality analysis explains predicates under which a failure happens and in their 
absence the failure would not happen under any other similar scenario.
Finding such causes provides engineers with tremendous insights to design more reliable systems, 
but it has been a long-standing and very challenging problem for various reasons, from defining a 
formal definition of causal effect of events to the high computational complexity of counterfactual 
reasoning. 

There is a wealth of research on causality analysis in the context of embedded 
and component-based systems from different 
perspectives~\cite{zl08,b24,ga14,gs20,cifg19,gss17,wg15,gmflx13,gmr10}.
Recently, there has been great interest in using temporal logics to reason about causality and \revision{explain} bugs~\cite{cdffhhms22,fk17,cffhms22,bffs23}.
In the CPS domain, using causality to repair AI-enabled controllers has recently gained 
interest~\cite{lcslr24,lrcsl23}.
However, these lines of work either focus on only modeling aspects of causality or do not 
address the problem of scalability in automated reasoning about causality, which inherently 
involves a \revision{combinatorial blow up to enumerate counterfactuals}.

\begin{tcolorbox}[title=Objectives]
\revision{This paper is concerned with the following problem. Given are (1) a formal operational 
description of a computing system $\trans$ (e.g., a transition system of a CPS), and (2) a logical 
predicate 
$\varphi_e$ that describes the effect (e.g., a safety failure) as input.
Our goal is to identify a predicate $\varphi_c$ that describes the cause of $\varphi_e$ happening 
(e.g., the earliest bad decision made by a controller). We note that $\varphi_e$ can be given by the 
user or can be found by using a verification or testing technique.
Hence, the way the effect is identified is irrelevant to the problem studied in this paper.}
\end{tcolorbox}

\revision{The first natural step is to formalize the definition of causality and in fact,} there are several 
interpretations of the meaning of causality.
In this paper, we are motivated by the notion of {\em actual causality} by Halpern and Pearl 
(\HP)~\cite{h16}, which focuses on the causal effect of particular events, rather than type causality, 
which attempts to make general statements about scientific and natural phenomena (e.g., smoking 
causes cancer).
Actual causality is a formalism to deal with {\em token-level} causality, which aims to find the causal 
effect of individual events (in our context, in embedded and CPS), as opposed to {\em type-level} 
causality, which intends to generalize the causal effect of types of events.

As we aim at analyzing executions of computing systems \revision{(e.g., models or data logs of a 
CPS)}, we first formalize causal models in (possibly infinite-state) {\em 
transition systems}, rather than the classic set of structural equations~\cite{h16}.
\revision{We show that formalizing the three conditions of actual causality yields a 
\revision{second-order logic} formula} of the form $\varphi_\mathsf{hp} \triangleq \exists 
\tau.\exists \tau'.\forall \tau''.\psi$, where $\tau$, $\tau'$, and $\tau''$ 
range over the set of executions of a transition system \revision{and $\psi$ stipulates the relation 
between actual and counterfactual worlds.}
More specifically:
	
	\begin{itemize} 
		\item \revision{The outermost existential quantifier in $\varphi_\mathsf{hp}$ intends to 
		establishes a relationship between the cause and the effect in an execution $\tau$ (known as 
		the {\bf AC1} necessity condition in the \HP framework~\cite{h16}). 
		That is, the cause and then the effect {\em actually} happen in $\tau$.}
\item \revision{The inner existential quantifier aims at refuting the causal effect relation in the 
counterfactual world (known as the {\bf AC2(a)} condition, stipulating the ``but-for" condition under 
contingencies).
That is, when the cause does not happen, the effect will not happen.}

\item \revision{The universal quantifier requires that if the cause happens in any execution 
$\tau''$ that is to $\tau$ (as far as the variables contributing to the cause and effect are concerned), 
then the effect should also happen (known as the {\bf AC2(b)} sufficiency condition).}
\end{itemize}
This formula exhibits a quantifier alternation and indeed, the problem of deciding actual causality in 
a causal model is known to be {\sf \small DP-complete}~\cite{achi17} in the size of the model, 
illustrating the computational complexity of the problem.
To deal with this complexity, we propose an effective method to formally reason about actual 
causality using decision procedures to solve satisfiability modulo theory (SMT).
Although there has been tremendous progress in developing efficient SMT solvers, they may not 
scale well when dealing with very large causal models or data logs. 
To tackle this problem, we introduce a novel technique 
based on {\em abstraction-refinement} that allows identifying causes within smaller abstract causal 
models.
This abstraction simplifies the model and attempts to view it from a higher level, while preserving the 
causal relations.

\begin{figure}[t]
    \centering
    \begin{minipage}[t]{0.45\linewidth}
        \centering
        \scalebox{.55}{
        \begin{tikzpicture}
  \draw[dashed] (0,0) ellipse (4.2cm and 2.7cm);
  \draw[ultra thick] (0,0) circle (3cm and 1.4cm);
  \draw [dashed](0,0) ellipse (2.2cm and .7cm);
  \node[font=\small,minimum width=1cm, align=center] at (0,0) {Under-Approximation $\undr{\trans}$ \\ ({\bf AC1, AC2(a)}): $\exists \tau.\exists \tau'$};
  \node[font=\small,minimum width=1cm, align=center] at (0,1) {Concrete model $\trans$};
  \node[font=\small,minimum width=1cm, align=center] at (0,2) {Over-Approximation $\hat{\trans}$ \\ ({\bf AC2(b): $\forall \tau''$})};
\end{tikzpicture}
        }
        \caption{Over/under-approximations of the concrete model and their relation to \HP conditions of the form $\exists\exists\forall$.}
        \label{fig:absref}
    \end{minipage}
    \hfill
    \begin{minipage}[t]{0.45\linewidth}
        \centering
        \scalebox{.6}{
        \begin{tikzpicture}[node distance=2.5cm, auto, ultra thick]
    \node[rectangle] (start) {Concrete TS ($\trans$)};
    \node[draw,rectangle, rounded corners=5pt, below = 0.5cm of start](under){under-approximation ($\check{\trans}$)};
    \node[rectangle, below of=under] (ac1ac2a) {$\mathbf{AC1}\land \mathbf{AC2(a)}$};
    
    \node[draw,rectangle,rounded corners=5pt, below = 0.8cm of ac1ac2a](over){over-approximation ($\ovr{\trans}$)};
    \node[rectangle,below left of = under] (ac1ac2a1) {{$\mathbf{AC1} \, \wedge \, \neg \mathbf{AC2(a)}$}};
    \node[rectangle, below right of=under] (ac1ac2a2) {$\neg \mathbf{AC1} \land \mathbf{AC2(a)}$};
    
    \node[rectangle, below =0.85cm of over] (ac2b) {$ \mathbf{AC2(b)}$};
    \node[rectangle, right of=ac2b] (ac2b1) {$\neg \mathbf{AC2(b)}$};
    
    \path[->] (under) edge (ac1ac2a)
              (under) edge (ac1ac2a1)
              (under) edge (ac1ac2a2);
    \path[->] (ac1ac2a) edge (over)
                (over) edge (ac2b)
              (over) edge (ac2b1);
    \draw[->] (start) to (under);
    \draw[->, bend right,dotted] (ac2b1) to (over);
    \draw[->, bend left,dotted] (ac1ac2a1) to (under);
    \draw[->, bend right,dotted] (ac1ac2a2) to (under);
    \node[left,yshift=-0.8cm,xshift=1.7cm,] at (over.east) {refine};
    \node[left,yshift=-1cm,xshift=1cm,] at (under.east) {refine};
    \node[left,yshift=-1cm,xshift=.2cm,] at (under.west) {refine};

\end{tikzpicture}
        }
        \caption{Overall idea of our algorithm -- Steps of abstraction-refinement approach.}
        \label{fig:alg_tree}
    \end{minipage}
    \vspace{-4mm}
\end{figure}
Although the idea of abstracting causal models in terms of structural equations has been studied in 
the literature~\cite{rwbmjgs17,beh19,bh19}, \revision{these works develop an exact simulation which 
may not exist or do not attempt to establish a relation between actual causes in the abstract and 
concrete causal models.}
Our technique incorporates two levels of abstraction to reason about actual causality  (i.e., formula 
$\varphi_\mathsf{hp} \triangleq \exists \tau.\exists \tau'.\forall \tau''.\psi$).
More specifically, our approach works as follows (see Figs.~\ref{fig:absref} and~\ref{fig:alg_tree}).
Given a concrete causal model $\trans$:

\begin{itemize}
	\item  We first compute an under-approximate model $\undr{\trans}$ of $\trans$.
	This model is used to find witnesses for conditions {\bf AC1} and {\bf AC2(a)} (i.e., the existential 
	quantifiers).
	If not successful, we refine $\undr{\trans}$ (e.g., by including states that are in $\trans$ and 
	not in $\undr{\trans}$) and try again.
	
	\item If the previous step succeeds, we compute an over-approximate model $\ovr{\trans}$ to 
	verify condition {\bf AC2(b)} for the universal quantifier.
	If successful, then an actual cause is identified and the algorithm terminates.
	Otherwise, we refine $\ovr{\trans}$ (e.g., by excluding states that are in $\ovr{\trans}$ and 
	not in $\trans$) and repeat the second step\footnote{Alternatively, one can return to the first step 
	and start from scratch.}.
\end{itemize}
We prove the correctness of our approach by showing that our algorithm is sound (but not 
necessarily complete).

We have implemented\footnote{Source code and all trace logs available at 
\url{https://github.com/TART-MSU/Causality_abs_refinement}}our approach using the 
Python programming language and utilized the SMT 
solver Z3~\cite{dmb08} and data analysis libraries~\cite{numpy,pandas} to construct our solver and 
abstraction technique. 
We conduct experiments on \revision{three} case studies to find the actual 
cause of
violations of safety in (1) a neural network controller for a mountain car~\cite{openaigym}, 
(2) a controller for a Lunar Lander obtained by reinforcement learning~\cite{openaigym}, and 
\revision{(3) an MPC controller for an F-16 autopilot 
simulator~\cite{ARCH18:Verification_Challenges_in_F_16}.}
Our experiments demonstrate the effectiveness of our abstraction-refinement technique by several 
orders of magnitude compared to the SMT-based approach for concrete causal models.

In summary, our contributions are the following.
We:

\begin{itemize}
	\item Formulate the classic \HP framework by transition systems and introduce \revision{an} 
	SMT-based 
	decision procedure to identify actual causes in a computing system;
	
	\item Introduce a technique based on abstraction-refinement to deal with scalability of formal reasoning about actual causality, and
	
	\item Conduct \revision{three} rigorous experimental evaluations on AI-enabled as well as non-AI 
	controllers in CPS.
\end{itemize} 
\revision{Our work is the first step in automating discovery of actual cause of failures, and our 
experiments show that we are able to identify the earliest bad decisions by controllers that lead \revision{to} violations 
of safety requirements.}

\paragraph*{Organization} The rest of the paper is organized as follows.
Section~\ref{sec:pre} presents the classic \HP framework for actual 
causality.
In Section~\ref{sec:smt}, we introduce our formulation of \HP for transition 
systems as well as a translation to an SMT-based decision procedure to 
identify actual causes.
Our abstraction-refinement technique is introduced in 
Section~\ref{sec:absref}.
We present our experimental evaluation in Section~\ref{sec:eval}.
Related work is discussed in Section~\ref{sec:related}.
Finally, we make concluding remarks and discuss future work in 
Section~\ref{sec:concl}.
\section{Preliminaries -- Actual Causality}
\label{sec:pre}

In this section, we present the notion of {\em actual causality} by 
Halpern and Pearl (\HP)~\cite{h16} \revision{as the baseline preliminary concept}.
 \revision{Since, the definition in~\cite{h16} is not a natural model of computation,  
in Section~\ref{sec:smt}, we will adapt the concepts in this section to transition systems and 
second-order logic formulas in order to reason about actual causality in computing systems.
We will consistently use the {\em Mountain Car} running example to explain the definitions and 
concepts throughout the paper. }

\subsection{Causal Models}

\vspace{1mm}
\begin{definition}
	\label{def:signature}
	A \emph{signature} $\sig$ is a tuple $(\exvars, \envars, \range)$, where $\exvars$ is set of 
	\emph{exogenous} variables (variables that represent factors outside the control of the model), 
	$\envars$ is a set of \emph{\ENDO} variables (variables whose values are ultimately determined 
	by the values of the \ENDO and exogenous variables).
	$\range$ is a function that associates with every variable $Y \in \exvars \cup \envars$ a nonempty set $\range(Y)$ of possible values for $Y$.~\qed
\end{definition}

Following Definition~\ref{def:signature}, a {\em state} is a valuation of a vector of variables $\vec{X} = (X_1, \ldots, X_n)$ in $\exvars \cup \envars$, where each variable $X \in \vec{X}$ is assigned a value from $\range(X)$. 

\vspace{1mm}
\begin{definition}
	\label{def:basic}
	A \emph{basic causal model} $\MOD$ is a pair $(\sig, \streq)$, where $\sig$ is a signature and $\streq_X$ defines a function that associates with each \ENDO variable $X$ a \emph{structural equation} $\streq_X$ that 
	%
	%
	maps $\range(\exvars \cup \envars - \{X\})$ to $\range(X)$, so $\streq_X$ determines the value of $X$, given the values of all the other variables in $\exvars \cup \envars$.~\qed
\end{definition}
It is important to highlight that \EXO variables cannot be linked to a function; thus, assigning values to \EXO variables, denoted as $\vec{u}$, is referred to as a \emph{context}.  


\vspace{1mm}
\begin{definition}
	An {\em intervention} entails setting the values of \ENDO\ variables, denoted as $\vec{X} \leftarrow \vec{x}$, and this notation signifies that the variables within set $\vec{X}$ are assigned values $\vec{x} = (x_1, \ldots, x_n)$.~\qed 
\end{definition}

The structural equations define what happens in the presence of interventions. 
Setting the value of some variables $\vec{X}$ to $\vec{x}$ in a causal model $\MOD = (\sig,\streq)$ 
results in a new causal model, denoted $\MOD_{\vec{X} \leftarrow \vec{x}}$, which is identical to 
$\MOD$, except that $\streq$ is replaced by $\streq^{\vec{X} \leftarrow \vec{x}}$: for each variable  
$Y \notin \vec{X}$, $\streq_Y^{\vec{X}\leftarrow\vec{x}} = \streq_Y$ while for each $X'$ in 
$\vec{X}$, the equation $\streq_{X'}$ is replaced by $X' = x'$. 
Thus, we define a {\em causal mode}l $\MOD$ by a tuple $(\sig, \streq, \intrv)$, where $(\sig, \streq)$ is a basic causal model (see Definition~\ref{def:basic}) and $\intrv$ is a set of \emph{allowed interventions}. 
Following~\cite{bh19}, the sets of ``allowed interventions'' ensure that the interventions can be appropriately limited to include only those that can be abstracted.

\begin{figure}[t]
    \centering
    \begin{subfigure}[b]{0.45\linewidth}
        \centering
        \resizebox{\linewidth}{!}{\begin{tikzpicture}

\draw[line width=0.5pt,-] (-3.5,-4) -- (3.5,-4) node[right] {$pos$};
\foreach \x/\label in {-3/-1.2,1/0,3/0.6}
    \draw (\x,-3.9) -- (\x,-4) node[below] {\label};
\draw[line width=1pt] (-3,0) to[out=-90, in=180] (1,-4);
\draw[line width=1pt] (1,-4) to[out=0, in=-90] (3,0);
\draw[line width=1pt] (-3,0) -- (-3.5,0);
\draw[line width=1pt] (3,0) -- (3.5,0);
\draw[line width=1pt, fill=red] (3,0.3) -- (3.4,0.45) -- (3,0.6) -- cycle;
\draw[line width=1pt] (3,0.3) -- (3,0);
\draw[line width=1pt, fill=black] (0.4,-3) rectangle (1.6,-3.52);
\draw[line width=1pt, fill=black] (0.6,-2.97) .. controls (0.8,-2.6) and (1.2,-2.6) .. (1.4,-2.97) -- cycle;
\draw[line width=1pt, fill=gray] (0.7,-3.75) circle (0.2);
\draw[line width=1pt, fill=gray] (1.3,-3.75) circle (0.2);
\draw[line width=1pt, fill=red] (0.4,-3.1) rectangle (0.5,-3.3);
\draw[line width=1pt, fill=yellow] (1.5,-3.1) rectangle (1.6,-3.3);
\end{tikzpicture}}
        \caption{} 
        \label{fig:car_M}
    \end{subfigure}
    \hfill
    \begin{subfigure}[b]{0.45\linewidth}
        \centering
        \resizebox{\linewidth}{!}{\begin{tikzpicture}[node distance=1, >=Stealth, every node/.style={circle, draw, minimum size=5em}]
    \node (init_pos) [circle, draw] {$\pos(t)$};
    \node (init_vel) [circle, draw, right=of init_pos] {$\vel(t)$};
    \node (g) [circle, draw, right=of init_vel] {$g$};
    \node (action) [circle, draw, below=of g] {$\action(t)$};
    \node (vel) [circle, draw, below=of init_vel] {$\vel(t+1)$};
    \node (pos) [circle, draw, below=of init_pos] {$\pos(t+1)$};

    \draw[->] (init_pos) -- (pos);
    \draw[->] (init_vel) -- (vel);
    \draw[->] (g) -- (vel);
    \draw[->] (action) -- (vel);
    \draw[->] (vel) -- (pos);
    
\end{tikzpicture}}
        \caption{}
        \label{fig:Car_graph}
    \end{subfigure}
    \caption{(a) Schematic of the mountain car example. (b) Graph illustrating the causal model and relationships between the variables at a snapshot in time $t$.}
    \label{fig:car_all}
    \vspace{-5mm}
\end{figure}
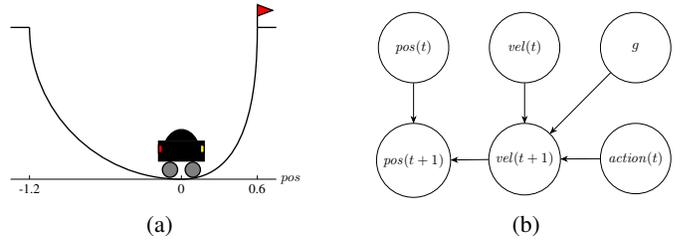

\begin{example}
	\label{ex:model}
Consider a car located in a valley and aiming to reach the top of a mountain (see 
Fig.~\ref{fig:car_M}). 
At each time step, the controller of the car determines whether to apply positive or negative acceleration to guide the car towards the mountain top. 
We define signature $\sig =(\exvars, \envars, \range)$ for this example as follows.
Let 
$$
\exvars = \{\pos(0), \vel(0), g\}
$$
be the set of exogenous variables, denoting the initial position, initial velocity, and the gravitational force on the car, respectively. 
Let 
$$
\envars = \{\pos(t), \vel(t), \action(t)\}
$$
be the set of endogenous variables, denoting the position, velocity, and the controller action, respectively, at each time step $t$, where $t \neq 0$. 
We also set:
\begin{align*}
\range(\pos(t)) = & [-1.2, 0.6]\\
\range(\vel(t)) = & [-0.07,0.07]\\
\range(\action(t)) = &\{-1, 0,1\}
\end{align*}  
where -1,0, and 1 are assigned as accelerate to the left, don't accelerate, and accelerate to the right, respectively, for all $t \geq 0$.
%

%


Now, we define the causal model $(\sig,\streq)$ based on the system dynamics for each $t > 0$ by 
structural equations:
\begin{align}
\label{eq:pos}
\streq_{\pos(t+1)} &= \streq_{\pos(t)} + \streq_{vel(t)} \\
\label{eq:vel}
\streq_{\vel(t+1)} &= \streq_{\vel(t)} + 0.001\streq_{\action(t)} - g.cos(3 \streq_{\pos(t)})
\end{align}
To illustrate the dependencies of the system, we can use a causal graph, as shown in Fig.~\ref{fig:Car_graph}. 
In this model, $\mathcal{M}_{\action(t)\leftarrow 1}$ denotes the model obtained by an intervention, 
where the $\action(t)$ is set to $1$ at time $t$ (for some $t > 0$).~\qed

\end{example}

\vspace{-2mm}
\subsection{Causal Formulas}

\label{sec:causal_form}

\revision{To precisely define actual causality, formal language is essential for articulating causal 
statements with clarity and rigor, in particular to formalize causes and effects. }
We use an extension of propositional logic, wherein primitive events take the form $\vec{X} = \vec{x}$, representing an \ENDO\ variable $\vec{X}$ and a possible value $\vec{x}$ for $\vec{X}$. 
The combination of primitive events is achieved through standard propositional connectives such as $\{\wedge, \vee, \neg\}$.
Thus, in this paper, we are only concerned with causal formulas that are state based (and not 
temporal).

	Given a signature $\sig =(\exvars, \envars, \range)$, a \emph{primitive event} is a formula of the 
	form $X =x$, for $X \in \envars$ and $x \in \range(X)$.
	A \emph{causal formula} (over $\sig$) is one of the form $\left[ Y_1\leftarrow y_1,\dots,Y_k\leftarrow y_k \right] \varphi$, where $\varphi$ is Boolean combination of primitive events, $Y_1,\dots,Y_k$ are distinct variables in $\envars$, and $y_i \in \range(Y_i)$.
	Such a formula is abbreviated as $[\vec{Y}\leftarrow \vec{y}]\varphi$. 
	The special case where $k = 0$ is abbreviated as $[ \space ]\varphi$ or, more often, just $\varphi$. 
	Intuitively, $[\vec{Y}\leftarrow \vec{y}]\varphi$ says that $\varphi$ would hold if $Y_i$ were set to $y_i$, for $i = 1,\dots, k$.

A causal formula $\psi$ is true or false in a causal model, given a context.
We use a pair $\MU$ consisting of a causal model $\MOD$ and context $\vec{u}$ as a \emph{causal 
setting}.
Hence, we write $\MU \models \psi$ if the causal formula $\psi$ is true in the causal setting $\MU$.
We are restricted to recursive models, where given a context, no cyclic dependencies exists.
In a recursive model, $\MU\ \models X = x $ if the value of $X$ is $x$ once we set the \EXO  variables to $\vec{u}$. 
Given a model $\MOD$, the model that describes the result of this intervention is $\MOD_{\vec{Y}\leftarrow \vec{y}}$.
Thus, $\MU\models [\vec{Y}\leftarrow y]\psi$ iff $(\MOD_{\vec{Y}\leftarrow \vec{y}},\vec{u}) \models \psi$.
Mathematical formalism serves to express the intuition precisely encapsulated within the formula 
$[\vec{Y}\leftarrow \vec{y}]\psi$ is true in a causal setting $\MU$ exactly if the formula $\psi$ is true 
in 
the model that results from the intervention, in the same context $\vec{u}$.

\begin{example}
\label{ex:setting}
Context  $\vec{u}$ in causal setting $(\MOD, \vec{u})$ in our example is determined by system inputs: initial velocity, initial position, and gravity:
$$
\vec{u} = \Big\{(\vel(0) \leftarrow 0.01), ( \pos(0) \leftarrow 0 ), ( g \leftarrow 0.0025 )\Big\}.
$$ 
where we defined $\MOD$ in Example~\ref{ex:model}.
To conduct causal analysis, the car at time $t = 0$ decides to set $\action(0) = 1$, but it fails to 
reach the goal. 
We defined causal formula to express failure as follows:
$$
\revision{\varphi_\fail \; \triangleq \; \Big(\pos(n) \neq 0.6\Big)},
$$
where $0.6$ is the flag position and $n$ is the last car state.~\qed

\end{example}

\vspace{-2mm}
\subsection{Actual Causality}


\vspace{1mm}
\begin{definition}
	\label{def:actualcause}
	$\vec{X} \leftarrow \vec{x}$ is an {\em actual cause} of $\varphi$ in causal setting $\MU$, if the following three conditions hold:
	
	\begin{itemize}
		\item \textbf{AC1.} $\MU \models [\vec{X} \leftarrow \vec{x}]\varphi$
		
		\item  \textbf{AC2(a).} There is a partition of $\mathcal{V}$ (set of \ENDO\ variables) into two disjoint 
		subsets $\vec{Z}$ and $\vec{W}$ (i.e, $\vec{Z}\cap\vec{W}=\emptyset$) with 
		$\vec{X}\subseteq\vec{Z}$ and a setting $\vec{x}'$ and $\vec{w}$ of the variables in $\vec{X}$ and 
		$\vec{W}$, respectively, such that:
		$$
		\MU\models [\vec{X}\leftarrow\vec{x}', \vec{W}\leftarrow 
		\vec{w}]\ \neg\varphi.
		$$
		
		\item  \textbf{AC2(b).} For all subsets $\vec{Z'}$ of $\vec{Z} - \vec{X}$, we have
		$$
		\MU\models [\vec{X} \leftarrow \vec{x}, \vec{W} \leftarrow \vec{w}, \vec{Z'} \leftarrow \vec{z^*}] \varphi
		$$
		where $\vec{z^*}$ denotes that variables in $\vec{Z'}$ are fixed at their values in the actual context.
		
		\item  \textbf{AC3.} $\vec{X}$  is minimal; no subset of $\vec{X}$ satisfies AC1 and AC2.~\qed
	\end{itemize} 
	
\end{definition}

Roughly speaking Definition~\ref{def:actualcause} expresses the following.
AC1 says that $\vec{X} = \vec{x}$ cannot be considered a cause of $\varphi$ unless both  $\vec{X} = \vec{x}$ and $\varphi$ actually happen.
AC2(a) says that the but-for condition holds under the contingency $\vec{W} = \vec{w}$.
Also, changing the value of some variable in $\vec{X}$ results in changing the value(s) of some variable(s) in $\vec{Z}$ (perhaps recursively), which finally results in the truth value of $\varphi$ changing.  
Finally, AC2(b) provides a sufficiency condition: if the variables in $\vec{X}$ and an arbitrary subset $\vec{Z} - \vec{X}$ of other variables on the causal path are held at their values in the actual context, then $\varphi$ holds even if $\vec{W}$ is set to $\vec{w}$ (the setting for $\vec{W}$ used in AC2(a)).
The types of events that the \HP definition allows as actual causes are ones of the form 
$
X_1 =  x_1\wedge \dots \wedge X_k = x_k
$, that is, conjunctions of primitive events; this is often abbreviated as $\vec{X}$.
\revision{In Section~\ref{sec:smt}, Example~\ref{ex:car_cause}, we will provide an example on how 
actual cause of formula $\varphi_\fail$ can be identified using our proposed technique.}

\section{SMT-based Discovery of Actual Causality}
\label{sec:smt}

In this section, we \revision{transform} the components of the \HP framework presented in 
Section~\ref{sec:pre} 
into transition systems and a second-order formula to express actual causality. \revision{Such a 
transition system can model the operational behavior of a system (e.g., a controller).
	Our technique can be agnostic to the details of the system and only take a set of execution 
	traces.}

Recall that a causal model $\MOD$ is of the form $(\sig, \streq, \intrv)$, where  $\sig = (\exvars, \envars, \range)$.
Also, recall that a state is a mapping from the variables in $\exvars \cup \envars$ to their respective domain of values.
We start with representing $\MOD$ with a set of traces obtained from a transition system that 
essentially describes how the state of all variables in $\exvars \cup \envars$ evolve over time by 
structural equations $\streq_X$, for every $X \in \envars$. 

\subsection{Transition Systems}
\label{sec:smt-ts}

\vspace{1mm}
\begin{definition}
	\label{def:transys}
A {\em transition system} $\trans$ corresponding to a causal model $\MOD$ is a tuple $\trans = (\States, \Trans, \Init, \Labels)$, where 

\begin{itemize}
\item $\States$ is the set of all possible states obtained from all possible valuations of variables in $\exvars \cup \envars$;

    \item $\Trans$ is a function mapping states in $2^{\States}$ to a state in $\States$ (recall that $\streq_X$ is a function);

\item $\Init \in \States$ is the initial state, and

\item $\Labels$ is a function mapping states in $2^\States$ to an atomic proposition from a set $\AP$ (e.g., given by causal formulas).~\qed 
\end{itemize}
\end{definition}

Following Definition~\ref{def:transys}, given a causal setting $(\MOD, \vec{u})$, the corresponding 
{\em causal transition system} is one that is acyclic and fixes $\sigma_0$ according to $\vec{u}$.
%
%
An intervention $\vec{X} \leftarrow \vec{x}$ is simply a set of transitions in $\Trans$ where in the target state $\vec{X} = \vec{x}$ holds, denoted by $\Trans_{\vec{X} \leftarrow\vec{x}}$.
We denote the set of all possible interventions in $\trans$ by $\intrv_\trans$.

\vspace{1mm}
\begin{definition}
	A {\em path} of a transition systems $\trans = (\States, \Trans, \Init, \Labels)$ is a sequence of 
	states of form $\sigma_0\sigma_1 \ldots$, where for all $i \geq 0$ (1) $\sigma_0= \Init$, and (2)  
	$(\sigma_i, \sigma_{i+1}) \in \Trans$. The {\em trace} corresponding to a path 
	$\sigma_0\sigma_1 \ldots$ is the sequence $\tau = \Labels(\sigma_0)\Labels(\sigma_1) 
	\ldots$.~\qed
\end{definition}
Let $\Traces$ denote the set of all traces of a transition system.

	   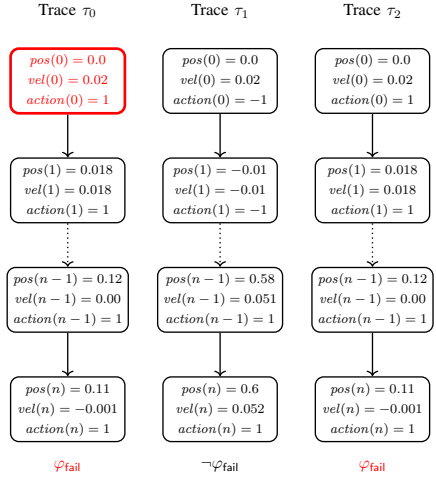
\begin{figure}[t]
	\centering
	\resizebox{0.65\linewidth}{!}{\begin{tikzpicture}[node distance=2.3cm, thick]
  \node[draw,ultra thick,color=red,rounded corners=5pt , minimum width=2.4cm, align=center] 
  (square1) 
  {\footnotesize$\pos(0)=0.0$\\\footnotesize$\vel(0)=0.02$ \\ \footnotesize$\action(0)=1$};
  \node[draw, rounded corners=5pt, minimum width=2.4cm, align=center, below of=square1] (square2) {\footnotesize$\pos(1)=0.018$\\\footnotesize$\vel(1)=0.018$ \\ \footnotesize$\action(1)=1$};
  \node[draw,rounded corners=5pt , minimum width=2.4cm, align=center, below of=square2] 
  (square3) {\footnotesize$\pos(n-1)=0.12$\\\footnotesize$\vel(n-1)=0.00$ \\ 
  \footnotesize$\action(n-1)=1$};
  \node[draw,rounded corners=5pt , minimum width=2.4cm, align=center, below of=square3] (square4) {\footnotesize$\pos(n)=0.11$\\\footnotesize$\vel(n)=-0.001$ \\ \footnotesize$\action(n)=1$};
  
  
  \draw[->] (square1) -- (square2);
  \draw[->, dotted] (square2) -- (square3);
  \draw[->] (square3) -- (square4);
  
  \node[above,yshift=0.5cm] at (square1.north) {Trace $\tau_0$};
  \node[yshift=-5mm] at (square4.south) {\textcolor{red}{$\varphi_{\fail}$}};
  
  \node[draw,rounded corners=5pt , minimum width=2.4cm, align=center,right of=square1, node distance=3.2cm] (square1_right) {\footnotesize$\pos(0)=0.0$\\\footnotesize$\vel(0)=0.02$ \\ \footnotesize$\action(0)=-1$};

  \node[draw, rounded corners=5pt, minimum width=2.4cm, align=center, below of=square1_right] (square2_right) {\footnotesize$\pos(1)=-0.01$\\\footnotesize$\vel(1)=-0.01$ \\ \footnotesize$\action(1)=-1$};
  \node[draw,rounded corners=5pt , minimum width=2.4cm, align=center, below of=square2_right] (square3_right) {\footnotesize$\pos(n-1)=0.58$\\\footnotesize$\vel(n-1)=0.051$ \\ \footnotesize$\action(n-1)=1$};
  \node[draw,rounded corners=5pt , minimum width=2.4cm, align=center, below of=square3_right] (square4_right) {\footnotesize$\pos(n)=0.6$\\\footnotesize$\vel(n)=0.052$ \\ \footnotesize$\action(n)=1$};
  
  \draw[->] (square1_right) -- (square2_right);
  \draw[->, dotted] (square2_right) -- (square3_right);
  \draw[->] (square3_right) -- (square4_right);
  
  \node[above,yshift=0.5cm] at (square1_right.north) {Trace $\tau_1$};
  \node[yshift=-5mm] at (square4_right.south) {{$\neg \varphi_{\fail}$}};

  \node[draw,rounded corners=5pt ,minimum width=2.4cm, align=center, right of=square1_right, node distance=3.2cm] (square1_right1) {\footnotesize$\pos(0)=0.0$\\\footnotesize$\vel(0)=0.02$ \\ \footnotesize$\action(0)=1$};
  \node[draw,rounded corners=5pt , minimum width=2.4cm, align=center, align=center, below of=square1_right1] (square2_right1) {\footnotesize$\pos(1)=0.018$\\\footnotesize$\vel(1)=0.018$ \\ \footnotesize$\action(1)=1$};
  \node[draw,rounded corners=5pt , minimum width=2.4cm, align=center, below of=square2_right1] (square3_right1) {\footnotesize$\pos(n-1)=0.12$\\\footnotesize$\vel(n-1)=0.00$ \\ \footnotesize$\action(n-1)=1$};
  \node[draw,rounded corners=5pt , minimum width=2.4cm, align=center, below of=square3_right1] (square4_right1) {\footnotesize$\pos(n)=0.11$\\\footnotesize$\vel(n)=-0.001$ \\ \footnotesize$\action(n)=1$};
  
  \draw[->] (square1_right1) -- (square2_right1);
  \draw[->, dotted] (square2_right1) -- (square3_right1);
  \draw[->] (square3_right1) -- (square4_right1);
  
  \node[above,yshift=0.5cm] at (square1_right1.north) {Trace $\tau_2$};
  \node[yshift=-5mm] at (square4_right1.south) {\textcolor{red}{$\varphi_{\fail}$}};

\end{tikzpicture}}
	
	\caption{Three traces for the mountain car example.} 
	\label{fig:tarce}
 \vspace{-4mm}
\end{figure}

\begin{figure*}[t]
		\small
	\begin{tcolorbox}
		\begin{align*}
		\textbf{AC1} \; \revision{\triangleq} \; &  \exists \tau \in \Traces.~\big(\tau \models \neg 
\varphi_\effect \U 
			(\varphi_\cause \land \F  \varphi_\effect)\big) 
			& (\varphi_c \text{ causes } \varphi_e \text{ in } \tau)\\ 
			\ \\
			\textbf{AC2(a)} \; \revision{\triangleq} \; & \exists \tau' \in \Traces. 	\big(\tau' \models \G(\neg 
			\varphi_c 
			~\land~ \neg 
			\varphi_e)\big) \land  \big(\tau \not \equiv_Z \tau'  \lor \tau \not \equiv_W \tau'\big) 
			& (\text{changes in the causal inhibits $\varphi_e$})\\
			\ \\
			\textbf{AC2(b)} \; \revision{\triangleq} \; & \forall \tau'' \in \Traces. 	\big((\tau'' \models (\neg 
			\varphi_e \U 
			\varphi_c)  
			\land  
			(\tau \equiv_Z \tau'' \land \tau \not \equiv_W \tau'')\big) \rightarrow \big(\tau'' \models \F 
			\varphi_e\big) 
			& \hspace{-3mm}(\text{in traces similar to $\tau$, $\varphi_c$ causes $\varphi_e$})
		\end{align*}
	\end{tcolorbox}
	\caption{\HP conditions adapted for causal transition systems.}
	\label{fig:HP}
 \vspace{-2mm}
\end{figure*}

\vspace{1mm}
\begin{example}	
\label{ex:car_cause}
Figure~\ref{fig:tarce} shows three traces $\tau_0$, $\tau_1$, and $\tau_2$ for our mountain car example for context $\vec{u} = (\pos(0) = 0.0, \vel(0) = 0.02)$.
In each step, the controller makes acceleration decisions.
Dotted transitions means the next state is not the immediate next time step.
The $n$th state is the last state of the trace.
As can be seen, traces $\tau_0$ and $\tau_2$ never reach position 0.6 (i.e., satisfying causal 
formula $\varphi_{\fail}$, meaning failing to reach the flag), while trace $\tau_1$ does (i.e., violating 
causal formula $\varphi_{\fail}$, meaning successfully reaching the flag).~\qed
	
\end{example}

We introduce three temporal operators to express the occurrence of causes and effects in traces:
%

\begin{itemize}

	\item For a state $\sigma$ and a proposition $p \in \AP$ \; iff \; $\sigma \models p$ \; 
	iff \; $p \in \Labels(\sigma)$.
	
	\item A trace $\tau = \tau_0\tau_1\ldots$ satisfies formula $\G p$ (read as `always $p$' and 
	denoted $\tau \models \G p$) \; iff \; \revision{$\forall i \geq 0. \tau_i \models p$}.
	
	\item A trace $\tau_0\tau_1\ldots$ satisfies formula $\F p$ (read as `eventually $p$' and 
	denoted $\tau \models \F p$) \; iff \; \revision{$\exists i \geq 0. \tau_i \models p$}.
	
	\item A trace $\tau_0\tau_1\ldots$ satisfies formula $p \U q$ (read as `$p$ until $q$' denoted $\tau 
	\models p \U q$) \, iff \, \revision{$\exists i \geq 0. (\tau_i \models q \wedge (\forall j < i. \tau_j 
	\models p))$}.
\end{itemize}

\subsection{SMT-based Formulation of Actual Causality}
\label{sec:smt-form-ac}

\revision{An SMT decision problem generally consists of two components: (1) the SMT instance (i.e., 
data elements such as variables, domains, functions, sets, etc), and (2) SMT constraints (i.e., 
first-order modulo theory involving quantified Boolean predicates with arithmetic).
In the context of our problem, the SMT instance consists of two parts (1) A set of elements for 
expressing a transition system $\trans$, or, a set of traces $\Traces$ (e.g., from a data log).
While the latter is simply a set of sequences of states (defined as a function from natural numbers to 
the full set of states), the former is specified by Boolean formulas from the unrolled transition 
system, similar to standard bounded model checking~\cite{cbrz01} without loops.}
(2) Our SMT \revision{model} formalize conditions {\bf AC1}, {\bf AC2(a)}, and {\bf AC2(b)} of 
Definition~\ref{def:actualcause} for transition systems (see Fig.~\ref{fig:HP}).
For simplification, we omit AC3 (minimality of cause), as it is not the most important constraint 
to reason about causal effect of events in a system.
Condition {\bf AC1} (in  Fig.~\ref{fig:HP}) means in the set $\Traces$, there exists at least one trace 
$\tau$, where effect $\varphi_e$ appears after cause $\varphi_c$ holds. 
Condition {\bf AC2(a)} requires the existence of one trace $\tau'$, where neither cause $\varphi_c$ nor effect $\varphi_e$ hold.
Additionally, trace $\tau'$ is not equivalent to trace $\tau$ (identified in {\bf AC1}) as far as variables in $W$ or $Z$ are concerned (i.e., the counterfactual worlds).
The remaining endogenous variables, the ones in $\vec{W}$, are off to the side, so to speak, but may still have an indirect effect on what happens.
Condition {\bf AC2(b)} requires that for all traces $\tau''$ that are similar to $\tau$ as far as causal 
variables in $Z$ are concerned, if cause $\varphi_c$ holds, then effect $\varphi_e$ hold some time 
in the future.

\revision{We clarify that while SMT solvers cannot directly encode temporal operators, one can 
easily encode them using the above expanded definitions by quantifiers over traces.}

 \begin{tcolorbox}[title=SMT Decision Problem]
	Given are (1) a causal transition system $(\trans, \vec{u})$ (or a set of traces $\Traces$ expressed 
	as a mapping from natural numbers to states), (2) a causal formula $\varphi_e$, (3) an 
	uninterpreted function representing $\varphi_c$, and (4) constraints \textbf{AC1}, \textbf{AC2(a)}, 
	and \textbf{AC2(b)}. The corresponding SMT instance is satisfiable \; iff \; the interpreted 
	$\varphi_c$ is an actual cause of $\varphi_e$ in $\trans$.
\end{tcolorbox}
%
 
 \begin{example}
 	\label{ex:car_explain}
  	We aim to identify the cause of the failure, denoted as $\varphi_{\fail}$, explained in our running 
  	example. 
 	For the sake of argument, let $X = \{\action(0) = 1\}$.
 	Since both $\pos$ and $\vel$ are dependent on the value of $\action$, they are part of $\vec{Z}$ or the causal path.
 	That is,
 	$$
 	\vec{Z} = \big\{\pos(t), \action(t), \vel(t) \mid t >1\big\}
 	$$
 	and, hence, $W = \{\}$ (since $\vec{W} \cap \vec{Z} = \emptyset$).
 	We now analyze the conditions of \HP:
 
\begin{itemize}
	\item  Starting with {\bf AC1}, one can instantiate $\tau$ (in Fig.~\ref{fig:HP}) with concrete trace $\tau_0$ in Fig.~\ref{fig:tarce}, indicating the satisfaction of the first condition.
	
	\item Moving to {\bf AC2(a)}, which involves counterfactual reasoning, when we change the actual 
	setting in {\bf AC1} to a counterfactual value $\action(0) = -1$, the car eventually reaches the goal 
	(i.e., 
	$\pos=0.6$). 
	This change allows the car to initiate a leftward movement, acquiring the necessary momentum to reach the flag, so flipping the failure $\varphi_\fail$ to success (i.e., $\neg \varphi_\fail$). 
	Consequently, {\bf AC2(a)} is satisfied by instantiating $\tau'$ (in Fig.~\ref{fig:HP}) with concrete trace $\tau_1$ (in Fig.~\ref{fig:tarce}).
	Also, notice that condition $\tau_1 \not \equiv_Z \tau_0$ is satisfied.
	
	\item Considering {\bf AC2(b)}, notice that trace $\tau_2$ is identical to $\tau_0$ as far as the 
	variables in $\vec{Z}$ are concerned (i.e., $\tau_0 \equiv_Z \tau_2$).
\revision{	Also, since $\vec{W} = \{\}$, changing variables in $\vec{W}$ while preserving the actual context results in an equivalent scenario to {\bf AC1}, which is already satisfied. 
	Thus, the only trace that can instantiate $\tau''$ (in Fig.~\ref{fig:HP}) is $\tau_2$, in which $\varphi_{\fail}$ becomes true.
    Note that the reason $\tau_0$ and $\tau_2$ are trace-equivalent is indeed due to the fact that $\Vec{W} = \{\}$.
	Hence, {\bf AC2(b)} hold.}
	
\end{itemize}

This means in this set of traces, $\action(0) = 1$ is the actual cause of failure for the car to reach the 
flag.~\qed
 	
 \end{example}

In the ideal world, one has to have {\em all} possible traces for combinatorial enumeration to evaluate {\bf AC2(b)}.
However, this is far from reality and most trace data logs (e.g., by some testing mechanisms, 
fuzzing, mutation testing, some automaton, etc) include only a subset of possibilities.
Our goal in this paper is to identify causal effects {\em within} a given set of traces.
%
%
\revision{Finally, as mentioned in the introduction, decision procedure for verification of actual 
causality is {\sf \small DP-complete}~\cite{achi17}, signifying the computation difficulty of automated 
reasoning 
about causality. This means our SMT-based problem is indeed dealing with a decision problem that 
is {\sf \small DP-complete}, setting the complexity of our SMT-based solution.}

\section{Abstraction-Refinement for Causal Models}
\label{sec:absref}

In this section, we propose our abstraction-refinement technique and its application in reasoning about actual causality, as presented in Section~\ref{sec:smt}.

\subsection{Overall Idea}

Generally speaking, the traditional abstraction approach to handle an existential quantifier is {\em under-approximation}, where we start from a subset of behaviors and attempt to instantiate the quantifier.
If successful, then the problem is solved.
Otherwise, we refine the abstraction by including addition behaviors and try again.
On the contrary, to handle universal quantifiers, the traditional abstraction approach is {\em over-approximation}, where we start from a subset of behaviors and attempt to verify universality.
If successful, then the problem is solved.
Otherwise, we need to ensure that the counterexample is not {\em spurious} (due to over-approximation). If it is, we refine the abstraction by excluding the counterexample and try again.

The overall idea of our technique is as follows (see Fig.~\ref{fig:alg_tree}).
Observe that the logic formula for actual causality is of the form $\exists\exists\forall$ (see 
Fig.~\ref{fig:HP}).
Given a transition system $\trans$ and causal formula $\varphi_e$ as the effect, we proceed as follows:


\begin{itemize}	
	
	\item {\bf Step 1.} \  Compute an under-approximation $\check{\trans}$ and an over-approximation $\hat{\trans}$.
	We first attempt to instantiate the existential quantifiers in {\bf AC1} and {\bf AC2(a)} in $\undr{\trans}$.
	If instantiating one of the quantifiers does not succeed, we refine $\undr{\trans}$ and repeat Step 1.

	\item {\bf Step 2.} When Step 1 succeeds, we compute $\ovr{\trans}$ and verify the universal quantifier in {\bf AC2(b)} for $\ovr{\trans}$.
	If successful, the witness to $\tau$ is a trace where the actual cause happens and we also obtain 
	a witness to $\varphi_c$ by the SMT solver.
	Otherwise, we can either refine $\ovr{\trans}$ and repeat Step 2 or refine $\undr{\trans}$ and return to Step 1.
\end{itemize}
	
We show that termination of these steps results in identifying an actual cause $\varphi_c$ in 
$\trans$ for $\varphi_e$.
This algorithm, however, may never terminate and, thus, our approach is sound but not complete.
\revision{We also remark that our heuristic based on abstraction-refinement is sound but not 
complete (e.g., similar to the CEGAR~\cite{cgjlv00} technique in model 
checking) to solve the general {\sf \small DP-complete} problem. The computation complexity of our 
solution, therefore, does not change.}

\subsection{Approximating Causal Transition Systems}

\label{sec:TS}

We first fix some notation.
For a {\em concrete} causal transition system $\trans = (\States, \Trans, \Init, \Labels)$ (the one given as input for causal reasoning), let us denote an over-approximate causal transition system by $\ovr{\trans} = (\ovr{\States}, \ovr{\Trans}, \ovr{\Init}, \ovr{\Labels})$ and an under-approximate causal transition system by  $\undr{\trans} = (\undr{\States}, \undr{\Trans}, \undr{\Init}, \undr{\Labels})$.
We denote the domain of endogenous (respectively, exogenous) variables of $\trans$ by $\range(\envars_\trans)$ (respectively, $\range(\exvars_\trans)$.

Given an over-approximate causal transition system $\ovr{\trans}$, we construct a sequence
$
\ovr{\trans}_0 \geq \ovr{\trans}_1 \geq \dots \ovr{\trans}_k 
$
of over-approximations, where (1) $\ovr{\trans}_k = \ovr{\trans}$, and $\ovr{\trans}_{i+1}$ is a 
refinement of $\ovr{\trans}_i$, for $0 \leq i < k$,  which we compute using \emph{counterexamples}.
A counterexample is a state of $\ovr{\States}_{i}$ that is not in $\States$. 
Over-approximation state mapping is a function which map states from $\trans$ to $\ovr\trans$; i.e., 
$
\FunOvrV:  2^{\States} \mapsto \ovr\States   
$. 
\begin{assumption}
	\label{assmp:Zvars}
In this paper, we only allow over-approximation state mappings $\FunOvrV$ that preserve the 
equality of traces as far as variables in $Z$ are concerned.
That is, for two concrete transitions $(\sigma_0, \sigma_1)$ and $(\sigma'_0, \sigma'_1)$, if (1) 
$\sigma_0 \equiv_Z \sigma'_0$, and (2) $\sigma_{1} \not \equiv_Z \sigma'_1$, then we have (1) 
$\sigma_0 \equiv_Z \FunOvrV(\sigma'_0)$, and (2) $\sigma_{1} \not \equiv_Z \FunOvrV(\sigma'_1)$.
 Otherwise, we will not be able to prove the soundness of Algorithm~\ref{alg:absref} with regard to 
 causal paths.
 We will elaborate more in the requirement in proof of Theorem~\ref{thrm:soundness}.
 We will also explain in Section~\ref{sec:eval}, how this assumption is ensured in our 
 implementation~\qed
 \end{assumption}
 	
%
%
%
We need an additional function:
$
\hintO: \intrv_\trans \mapsto \intrv_{\ovr{\trans}}
$
which maps concrete interventions to over-approximation interventions.

\vspace{1mm}
\begin{definition}
    	\label{def:restrict}
Given a subset of endogenous variables in $\States$, called $\vec{X}$, and $\vec{x} \in 2^{\States}$, let 
$$
\Rst(\States, \vec{x}) = \{ \vec{v} \in 2^{\States} : \vec{x} \text{~is the restriction of~} \vec{v} \text{~to~} \vec{X}\}.~\blacksquare
$$
	
\end{definition}
This definition carries to a transition system $\trans = (\States,\Init, \Trans, \Labels)$ in a straightforward fashion as follows.
The restriction of a set of values $\vec{x}$ on $\States$ is a subset $\States\rst_{\vec{x}} \subseteq 
\States$ restricted to those states, where $\Vec{X} = \vec{x}$.
The set of restricted transitions is obviously those start and end in states in $\States\rst_{\vec{x}}$.

We now explain how we compute the above functions.
Given $\FunOvrV$, we define: 
$
\hintO(\Trans_{\Vec{X} \leftarrow \vec{x}}) = \ovr{\Trans}_{\Vec{Y} \leftarrow \vec{y}}
$ 
if (1) $\vec{y} \in 2^{\ovr\States}$, and (2) 
$\FunOvrV(\Rst(\States\rst_{\vec{x}}))=\Rst(\ovr{\States}\rst_{\vec{y}})$. 
Hence for every intervention in $\Trans_{\Vec{X} \leftarrow \vec{x}}$, there is only one intervention in $\ovr{\Trans}_{\Vec{Y} \leftarrow \vec{y}}$.
If such a $\vec{Y}$ and $\vec{y}$ do not exist, we take $\hintO(\Trans_{\Vec{X} \leftarrow \vec{x}})$ to be {\em undefined}. 
Let  $\intrv_\trans^{\FunOvrV}$ be the set of interventions for which $\hintO$ is defined, and let $\intrv_{\ovr{\trans}} = \hintO(\intrv_\trans^{\FunOvrV})$.

%
	
%

Based on this definition, it becomes evident that not all interventions in $\intrv_\trans$ will have corresponding mappings in $\intrv_{\ovr{\trans}}$  or $\intrv_{\undr{\trans}}$. 
This is due to the fact that $\FunUndV$ and $\FunOvrV$ may aggregate states, resulting in some 
$\intrv_\trans$ representing only partial interventions on $\intrv_{\undr{\trans}}$ or 
$\intrv_{\undr{\trans}}$.
In this context, the introduction of a notion termed allowed intervention becomes crucial.
This notion is essential as certain interventions in the abstract model may lack definition or relevance in a well-defined concrete model. 
Consequently, within this framework of definitions, the translation of interventions is not universal; rather, only essential interventions that can be meaningfully mapped are considered. 
%

We follow a similar but simpler procedure for under-approximations.
Given an under-approximate causal transition system $\undr{\trans}$, we construct a sequence
$
\undr{\trans}_0 \leq \undr{\trans}_1 \leq \dots \undr{\trans}_k 
$
of under-approximations, where (1) $\undr{\trans}_k = \undr{\trans}$, and $\undr{\trans}_{i+1}$ is a 
refinement of $\undr{\trans}_i$, for $0 \leq i < k$,  which we compute using 
\emph{counterexamples}.
A counterexample is a state of $\States$ that is not in $\undr{\States}_{i}$.
In this paper, since we begin causal analysis from a trace log $\Trans$, we compute an 
under-approximation by a subset of the input set of traces.
That is, $\FunUndV(\Traces)  \subseteq \Traces$.
%
%

\subsection{Detailed Description of the Algorithm}
\label{sec:detalgo}

The input to Algorithm~\ref{alg:absref} is a concrete transition systems $\trans$ (more specifically, 
its trace set) and a causal formula $\varphi_e$. 
Also, $\alpha$ and $\beta$ and are parameters used in computing $\FunOvrV$ and $\FunUndV$, respectively. $\alpha$ indicates the subset size of $\FunUndV$ and $\beta$ is a threshold to compute Euclidean distance of states for over-approximation.
We are restricted to a set of allowed interventions $\intrv_\trans^{\FunOvrV}$. 
Our objective is to identify states of $\trans$, where causal formula $\varphi_c$ holds as an actual 
cause in the trace $\tau = \undr\Labels(\sigma_0)\undr\Labels(\sigma_1) \ldots$.

Line~\ref{line:init1} initializes the under-approximation $\undr{\trans}$, with 
parameter $\alpha$ indicating the number of traces to use and map in $\FunUndV$ function.
In lines~\ref{line:SMT1} -- \ref{line:underagain}, the algorithm computes whether the SMT query 
returns $\varphi_c$ as the cause for effect $\varphi_e$ in the current under-approximation and 
over-approximation. 
Specifically, in line~\ref{line:SMT1}, the SMT function receives $\undr{\trans}$ as the 
under-approximation and constraints of {\bf AC1} and {\bf AC2(a)} specified in Fig.~\ref{fig:HP}, and 
it returns the result $\varphi_c$ as the cause.
The SMT solver also returns a witness trace $\tau \in \undr{\Traces}$.
In line~\ref{line:underagain}, if the result of SMT query in line~\ref{line:SMT1} is unsatisfiability, then 
the algorithm chooses more traces $\Traces$ by increasing $\alpha$. 
Indeed, lines~\ref{line:inalpha} and~\ref{line:underagain} establish the refinement for the 
under-approximate model.
%

	
\begin{algorithm}[t]
\footnotesize
	\caption{Finding actual cause of $\varphi_e$ in $\trans$ }
		   \label{alg:absref}
	\KwIn{$\trans = (\States, \Trans, \Init, \Labels)$, causal formula $\varphi_e$, allowed interventions 
	$\intrv_\trans^{\FunUndV}$, $\alpha =[0,1]$, $\beta \geq 0$}
	\KwOut{Causal formula $\varphi_c$}
	$\undr{\Traces} \leftarrow \FunUndV({\Traces})$ using $\alpha$\; \label{line:init1}
	\While{true}{
		$\{\varphi_c, \undr{\tau}, \undr{\tau}'\}  \leftarrow$ SMT($\undr{\Traces}$, $\mathbf{AC1} \land 
		\mathbf{AC2(a)}$)\; \label{line:SMT1}
		\If {$\neg \varphi_c$} {
    $\ovr{\trans} \leftarrow \FunOvrV(\trans)$ using $\beta$ and $\varphi_c$\; \label{line:init2}
			\While{true}{
				$\text{result} \leftarrow$ SMT($\ovr{\trans}, \varphi_c$, {\bf AC2(b)})\; \label{line:SMT2}
				\eIf{$\text{result}$}{
					\textbf{return} $\varphi_c$ \;
     			}
				{$\ovr{\trans} \leftarrow \ovr{\Refine}(\ovr{\trans}, \States - 
				\ovr{\States},\intrv_\trans^{\FunUndV})$\; \label{line:refine}} 
			}
		}
		Increase $\alpha$\;\label{line:inalpha}
		$\undr{\Traces} \leftarrow \FunUndV({\Traces})$ using $\alpha$\; \label{line:underagain}
	}
\end{algorithm}

If a cause $\varphi_c$ is identified by satisfying {\bf AC1} and {\bf AC2(a)}, we use this cause to 
initialize over-approximation in line~\ref{line:init2} (to ensure Assumption~\ref{assmp:Zvars}), where 
we include all original states as well as potentially unreachable states by creating an abstract 
representation by function $\FunOvrV$, such that all states in $\trans$ map to $\ovr\trans$, and also 
similar states are merged into a single abstracted state in $\ovr\trans$.
The distance threshold for merging states is controlled by the parameter $\beta$. 
If the distance between any pair of states is less than $\beta$, those states will be merged.
Consequently, a smaller $\beta$ results in a larger number of abstract states, while a larger $\beta$ 
leads to a smaller number of abstract states.

In lines \ref{line:SMT2} -- \ref{line:refine}, we focus on verifying {\bf AC2(b)} using the 
over-approximation. 
In line \ref{line:SMT2}, the SMT query takes $\ovr{\trans}$ as the current over-approximate model 
and $\varphi_c$ as output from line~\ref{line:SMT1}. 
It then examines whether all traces for which $\varphi_c$ and $\varphi_e$ hold can be modified by changing states such that $\varphi_e$ still holds. 
If the SMT solver returns SAT, then $\varphi_c$ is returned as the actual cause, where $\varphi_c$ 
is a Boolean expression on the atomic propositions related to states in a specific trace. 
If the result is not SAT, in line~\ref{line:refine}, we use counterexample(s) in $\States - 
\ovr{\States}$, allowed interventions identified by $\hintO(\intrv_{\trans})$. 
These counterexamples are then eliminated by $\Refine$, and the resulting model is assigned to the 
new $\ovr{\trans}$.
%
%
We emphasize that in the refinement step for over-approximation (line~\ref{line:refine}), it is crucial 
to consider restricted interventions, denoted as $\intrv_\trans^{\FunUndV}$.
%
 This consideration is necessary because, in a concrete model, certain interventions may not be 
directly mapped to their counterparts in the over-approximated model.
Consequently, the refinement process must incorporate $\intrv_\trans^{\FunUndV}$ as an 
essential input, utilizing it effectively during the mapping process to ensure consistency of model 
translation.

\subsection{Correctness}

In this section, we formally prove the soundness of Algorithm~\ref{alg:absref}.

\vspace{1mm}
\begin{theorem}
	\label{thrm:soundness}
	Let $\trans$ be a concrete causal transition system and $\varphi_c$ and $\varphi_e$ be two 
	causal formulas.
	If $\varphi_c$ is an actual cause of $\varphi_e$ identified by Algorithm~\ref{alg:absref} (for 
	$\ovr{\trans}$ and $\undr{\Traces}$), then $\varphi_c$ is an actual cause of $\varphi_e$ in 
	$\trans$.
	
\end{theorem}

\begin{IEEEproof}
Let  $\ovr{\trans}$ and $\undr{\trans}$ be the over- and under-approximate models of $\trans$.
Formally, we need to prove the following.
{\bf Assumption:} Let us assume that:

	\begin{itemize}
		\item {\bf AC1} holds in $\undr{\Traces}$. That is:
		 $$\exists \undr{\tau} \in \undr{\Traces}.~\big(\undr{\tau} \models \neg \varphi_\effect \U 
		 (\varphi_\cause \land \F  \varphi_\effect)\big) 
		 $$
		 
		\item {\bf AC2(a)} holds in  $\undr{\Traces}$. That is:
		$$
		\exists \undr{\tau}' \in \undr{\Traces}. 	\big(\undr{\tau}' \models \G(\neg \varphi_c ~\land~ \neg 
		\varphi_e)\big) \land  \big(\undr{\tau} \not \equiv_Z \undr{\tau}'  \, \lor \, \undr{\tau} \not \equiv_W 
		\undr{\tau}'\big) 
		$$
		
		\item {\bf AC2(b)} holds in $\ovr{\trans}$. That is, 
		\begin{align*}
		\forall \ovr{\tau}'' \in \ovr{\Traces}. &	\big((\tau'' \models (\neg \varphi_e \U \varphi_c) \; \land  \\
		& (\ovr{\tau} \equiv_Z \ovr{\tau}'' \land \ovr{\tau} \not \equiv_W \ovr{\tau}'')\big) \rightarrow 
		\big(\ovr{\tau}'' \models \F \varphi_e\big) 
		\end{align*}
	\end{itemize}
{\bf Goal:} We should prove that:
	
	\begin{itemize}
	\item {\bf AC1} holds in $\trans$. That is:
	$$\exists \tau \in {\Traces}.~\big(\tau \models \neg  \varphi_\effect \U (\varphi_\cause \land \F  
	\varphi_\effect)\big) 
	$$
	
	\item {\bf AC2(a)} holds in  ${\trans}$. That is:
	$$
	\exists \tau' \in {\Traces}. 	\big(\tau' \models \G(\neg \varphi_c ~\land~ \neg \varphi_e)\big) \land  
	\big(\tau \not \equiv_Z \tau'  \lor \tau \not \equiv_W \tau'\big) 
	$$
	
	\item {\bf AC2(b)} holds in ${\trans}$. That is, 
	\begin{align*}
		\forall \tau'' \in {\Traces}. &	\big((\tau'' \models (\neg \varphi_e \U \varphi_c) \; \land  \\
		& (\tau \equiv_Z \tau'' \land \tau \not \equiv_W \tau'')\big) \rightarrow \big(\tau'' \models \F \varphi_e\big) 
	\end{align*}
\end{itemize}

To this end, we proceed as follows.
	
	
 \begin{itemize}

        \item We show that $\tau$ (in our goal) can be simply instantiated by the witness $\undr{\tau}$ 
        (in the assumption).
        In our first assumption, let $\undr{\tau} \in \undr\Traces$ be the trace returned in 
        line~\ref{line:SMT1} that satisfies {\bf AC1}.
        That is, effect $\varphi_e$ occurs after cause $\varphi_c$ in $\undr{\tau}$. So, we have,
        $\undr{\tau} = \undr{\Labels}(\sigma_{0})\undr{\Labels}(\sigma_{1})\dots $.
        There exists $i$ and $j$ where $i<j$, such that $\undr{\Labels}(\sigma_{i}) \models \varphi_c$ 
        and $\undr{\Labels}(\sigma_{j}) \models \varphi_e$.
        As described earlier, $\undr\Traces \subseteq \Traces$, so if there is a trace $\undr{\tau}$ in 
        $\undr\Traces$ that satisfies {\bf AC1}, and since all traces in $\undr\Traces$ must be inside 
        $\Traces$, the goal item is proved.

        \item We show that $\tau'$ (in our goal) can be simply instantiated by the witness $\undr{\tau'}$ 
        (in the assumption).
        In the second assumption, we assumed that there exists a trace $\tau' \in \undr\Traces$ that 
        satisfies {\bf AC2(a)}, which states that neither the cause $\varphi_c$ nor the effect 
        $\varphi_e$ hold in $\undr{\tau}'$ of the form, $\tau' = \undr{\Labels}(\sigma'_{0})\undr{\Labels}(\sigma'_{1})\dots$.
        
        That is, for all $i$ and $j$, $\undr{\Labels}(\sigma'_{i}) \models \neg \varphi_c$ and 
        $\undr{\Labels}(\sigma'_{j}) \models \neg \varphi_e$. 
        Since $\undr\Traces \subseteq \Traces$, if there is a trace $\undr{\tau}' \in \undr\Traces$ that 
        satisfies {\bf AC2(a)} the second goal is also proved.

        \item We start with the assumption that all traces $\ovr{\tau}'' \in \ovr\Traces$ satisfy 
        {\bf AC2(a)} (i.e., if agreeing with trace $\undr{\tau}$ on variables in $Z$ and disagreeing 
        with $\undr{\tau}$ with respect to variables $W$ the they satisfy the causal relation between 
        $\varphi_c$ and $\varphi_e$). 
        We prove this part by contradiction (i.e., if {\bf AC2(b)} does not hold in the concrete model 
        $\trans$, then it cannot hold in the over-approximate model $\ovr{\trans}$.
        Formally, the following formula should hold:
       	\begin{align*}
        	\exists \tau'' \in {\Traces}. &	\big((\tau'' \not \models (\neg \varphi_e \U \varphi_c) \; \vee  \\
        	& (\tau \not \equiv_Z \tau'' \vee \tau \equiv_W \tau'')\big) \wedge \big(\tau'' \models \F 
        	\varphi_e\big). 
        \end{align*}
       We distinguish two cases based on the two sides of the disjunction.:
       \begin{itemize}
       	\item Let us imagine $\tau'' \not \models (\neg \varphi_e \U \varphi_c)$, where $\tau'' = 
       	{\Labels}(\sigma_{0})\Labels(\sigma_{1})\dots $. If we compute $\ovr{\tau}'' = 
       	{\Labels}(\ovr{\sigma}_{0}){\Labels}(\ovr{\sigma}_{1})$.
       	Since over-approximation cannot change the satisfaction of formulas in abstract states, the 
       	order of truthfulness of formulas in states will not change.
       	Hence, we have $\ovr{\tau}'' \not \models (\neg \varphi_e \U \varphi_c)$, which contracts 
       the assumption of the theorem (i.e., {\bf AC2(b)} for the abstract models).
       
       \item Let us imagine $(\tau \not \equiv_Z \tau'' \vee \tau \equiv_W \tau'') \wedge \big(\tau'' 
       \models \F \varphi_e\big)$.
       Since $\tau'' \models \F \varphi_e$, we also have $\ovr{\tau}'' \models \F \varphi_e$.
       Again, we distinguish two cases:
       
       \begin{itemize}
       	\item If $\tau \equiv_W \tau''$ then it trivially holds that $\tau \equiv_W \ovr{\tau}''$ (since $\tau 
       	= \undr{\tau}$ and function $\FunOvrV$ preserves state valuations).
       	This case also reaches a contradiction, as {\bf AC2(b)} holds in the abstract model.
       	\item Finally, let us imagine $\tau \not \equiv_Z \tau''$. This will imply that $\tau \not \equiv_Z 
       	\ovr{\tau}''$, by Assumption~\ref{assmp:Zvars}, which contradicts the assumption of the 
       	theorem.
       \end{itemize}
       This case analysis shows that we reach a contradiction in all cases and, hence, the following 
       holds:
       	\begin{align*}
       	\forall \tau'' \in {\Traces}. &	\big((\tau'' \models (\neg \varphi_e \U \varphi_c) \; \land  \\
       	& (\tau \equiv_Z \tau'' \land \tau \not \equiv_W \tau'')\big) \rightarrow \big(\tau'' \models \F 
       	\varphi_e\big) 
       \end{align*}
       \end{itemize} 
This concludes the proof.
        %
 
\color{blue}

\end{itemize}

\end{IEEEproof}	

\section{Experimental Evaluation}
\label{sec:eval}

This section first provides an overview of the implementation details of the algorithm proposed in 
Section~\ref{sec:detalgo}.
\revision{We also evaluate our technique on three case studies: (1) Mountain Car, and (2) Lunar Lander environments from {\sf \small OpenAI Gym}~\cite{openaigym} -- commonly used evaluation benchmarks for learning-enabled CPS, and (3) an F-16 autopilot simulator~\cite{ARCH18:Verification_Challenges_in_F_16} that uses an MPC controller.}

\subsection{Implementation}
\label{sec:impl}
\revision{To identify the actual cause of failures in our studies, we need to generate traces consisting of those that do not violate safety and those that do violate safety.
We need successful traces to find counterfactuals for failure scenarios, where the same conditions lead to success through different decisions. 
}

\revision{In our experiments, we use 47 networks for the Mountain Car experiment~\cite{ivanov2018verisig,lrcsl23} and generate over 570 neural networks, trained with Deep Reinforcement Learning~\cite{dqn}, for the Lunar Lander case study.
%
%
Consequently, the success rates of traces in satisfying $\varphi_e$ used in our experiments were 17\% and 11\% for the case studies in Sections~\ref{sec:car} and~\ref{sec:lunar}, respectively. In the case study in Section~\ref{sec:f16}, the success rates were 21\% and 33\% for the first and second scenarios, respectively.
%
%
This is because the non-AI MPC controller typically makes better decisions than the AI controller.}

We have implemented Algorithm~\ref{alg:absref} using the Python programming language. 
\revision{Algorithm~\ref{alg:absref} is implemented through two approaches. First, the Z3 SMT 
solver~\cite{dmb08}, and secondly (for non-symbolic cases), by employing a search method to find 
traces in datasets that meet the \HP conditions. For instance, if we find a trace that leads to failure, 
we take this sample and search for other traces with the same features, except for the decision that 
caused the failure in the original trace. To accomplish this, we utilize built-in data science search 
algorithms in~\cite{numpy,pandas}.
} In fact, in our case studies, we are dealing with large-sized data rather than symbolic properties. 
\revision{Therefore, in Section~\ref{sec:analysis}, we will demonstrate that searching through the dataset is more efficient compared to Z3. While Z3 is primarily employed for its robust capabilities in theorem proving and constraint solving, it is not as effective for finding traces in a large set of already generated traces that meet certain conditions.
}

For execution of Algorithm~\ref{alg:absref}, specific strategies are adopted in refinement of the 
under- and especially over-approximate (function $\Refine$ in line~\ref{line:refine}) models in cases 
of unsatisfiability.
In the under-approximation model, a parameter $\alpha$ is utilized to incorporate additional traces. 
This parameter can be progressively increased to obtain more traces, thereby refining the under-approximation.
In the over-approximation model, a parameter $\beta$ is used within the mapping function to dictate the threshold for the distance between states. 
When the distance between a group of states is less than this threshold, they are merged into a single state to simplify the model. 
Moreover, in refining the over-approximate model, the algorithm checks for the existence of 
counterexample states that violate the over-approximation. 
If such states are identified, they are removed from the model to ensure its accuracy~\cite{cgjlv00}.
%

Assumption~\ref{assmp:Zvars} for both our case studies is implemented as follows that in the over-approximation function, it is crucial not to merge states that transition to different outcomes. 
For instance, in the Mountain Car example, if there are two traces that differ only in their actions but have the same position and velocity, and the under-approximation model identifies that action might be a possible cause of failure, these states cannot be merged in the over-approximation model. 
This is because merging them would obscure the distinction between a trace leading to failure and another leading to success.

\subsection{Experimental Settings}

All of our experiments were conducted on a single core of the Apple M2 Pro CPU, which features a 
10-core architecture and operates @3.7GHz. 
Given a set of collected traces, we applied our techniques in four different modes to identify the 
cause of failure (safety violations): 

\begin{itemize}

	\item \textbf{Only\_Z3} is the implementation, where we only use the SMT solver Z3 to discover 
	actual 
	causality (the technique proposed in Section~\ref{sec:smt}).

	\item  \textbf{Abs\_Z3} is the implementation, where Algorithm~\ref{alg:absref} uses Z3.

	\item  \textbf{Only\_DA} is the implementation, where we only use the search algorithms 
	in~\cite{pandas,numpy} in lieu of an SMT solver.

	\item  \textbf{Abs\_DA} is the implementation, where Algorithm~\ref{alg:absref} uses the search 
	algorithms in~\cite{pandas,numpy} in lieu of an SMT solver.

\end{itemize}
%

\subsection{Case Study 1: Mountain Car}
\label{sec:car}

Out first case study is the continuation of our running example.
In Fig.~\ref{fig:car_M}, the car is initially positioned in the valley between two mountains with the 
objective being to navigate it to the peak of the right mountain before a set deadline.
The system incorporates three variables in accordance with Equations \ref{eq:pos} and \ref{eq:vel}, specifying the domain for each variable as $\pos(t) \in [-1.2, 0.6]$, $\vel(t) \in [-0.07, 0.07]$, and $\action(t) \in [-1,1]$. 
Here, $\action$ represents a learning-based function $f$, implemented using various pre-trained neural networks of different dimensions:
$$\action(t)= f(\pos(t),\vel(t))$$
The car's mission is to achieve $\pos(t)=0.6$ before the time limit of $t=100$ episodes. 
Our study explores various initial settings for $\pos(0)$, $\vel(0)$, and the function $f$ to find the 
cause of the vehicle's failure to reach its target.
In our study, we began by collecting data by assigning different initial values to the variables 
$\pos(0)$ and $\vel(0)$, which were treated as external (\EXO) variables. \revision{We also} utilized 
various combinations of pre-trained neural networks as the decision-making mechanism for 
acceleration.
The action controller in the Mountain Car scenario employs a 
neural network characterized by a rectangular architecture with varied dimensions. 
The \textsf{\small sigmoid} function serves as the activation mechanism for both the input and 
hidden layers, 
whereas the \textsf{\small Tanh} function is utilized for the output layer. 
This approach represents a modification of the methodology detailed 
in~\cite{ivanov2018verisig,lrcsl23}. 

By executing multiple initial valuations with distinct neural networks, we generated a substantial set of traces, each indicating whether the car reached its destination within 100 episodes.

\subsection{Case Study 2: Lunar Lander}

\label{sec:lunar}

	

\revision{In this case study, the space lander is initially positioned at a certain altitude from the ground, aiming to land on the designated landing pad.}
The landing pad is always located at $(0\pm \epsilon, 0 \pm\epsilon)$.
In the Lunar Lander system, there are eight variables (e.g., $x$ and 
$y$ coordinates, velocity, angular velocity, angle, etc), which are 
intrinsic to the system, and an additional seven variables that are configured to represent the 
environment (e.g., wind, gravity, turbulence power, etc).
For a comprehensive overview of this case study, refer to~\cite{openaigym}. 
\revision{The exogenous variables we consider are four different values for wind: $\{0, 5, 10, 15\}$, three values for gravity: $\{-8, -10, -12\}$, and three values for turbulence power: $\{0.8, 1.5, 2\}$.}
Moreover, in our experiment, we focus on a subset of the endogenous variables, specifically 
$\pos_x(t)$, $\pos_y(t)$, $\vel_x(t)$, $\vel_y(t)$, and $\action(t)$. 
These variables correspond to the horizontal and vertical positions, horizontal and vertical velocities, 
and the action controlling the engines of the lander, respectively.

In our model, $\action$ denotes a learning-based function $f$, which is implemented using various 
pre-trained neural networks with different dimensions:
$$
\action(t) = f(\pos_x(t), \pos_y(t), \vel_x(t), \vel_y(t)).
$$
In the Lunar Lander environment, there are four discrete actions available for controlling the lander, 
denoted as $\action = \{0, 1, 2, 3\}$:

\begin{itemize}
    \item $0$: Do nothing;
    \item $1$: Fire the left orientation engine;
    \item $2$: Fire the main engine, and
    \item $3$: Fire the right orientation engine
\end{itemize}
The action controller designed for the Lunar Lander experiment is based on deep reinforcement 
learning principles, as explored in~\cite{dqn}. 
The neural networks employed in this experiment are rectangular in shape and utilize the Rectified Linear Unit (ReLU) activation function to introduce non-linearity and enhance the learning capability of the model.
\revision{We performed multiple simulations with varying initial values 
for $\pos_x(0)$, $\pos_y(0)$, $\vel_x(0)$, $\vel_y(0)$, $\wind$, $\gravity$, and $\torbu$ each paired with different 
neural networks.}
This procedure produced a substantial set of traces, with each trace indicating whether the lander 
successfully landed on the landing pad within the time frame of $t < 500$ episodes or not.

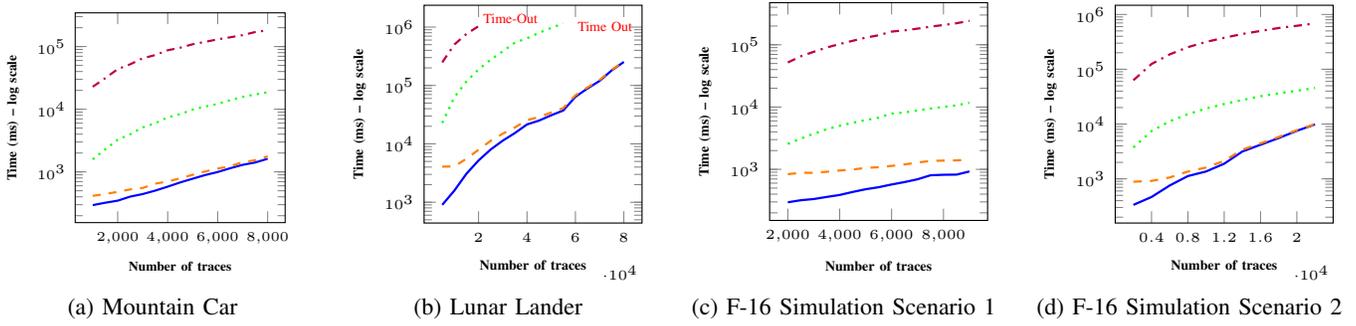
\begin{figure*}[t]
	\centering
	\begin{subfigure}[t]{0.24\linewidth}
		\centering
		\resizebox{\linewidth}{!}{ 

\begin{tikzpicture}
\begin{axis}[
    xlabel={\tiny \bf Number of traces},
    ylabel={\tiny \bf Time (ms) -- log scale},
    width=1\textwidth,
    height=1\textwidth,
    legend style={at={(0.8,-0.01)},anchor=south,draw=none, font=\tiny},
    ymode=log, 
    xtick={0,2000,4000,6000,8000},
    xticklabel style={font=\tiny},
    yticklabel style={font=\tiny}
]
\addplot[thick,blue,solid]  table [x=X, y=Y1, col sep=comma] {images/car.csv};\label{line:abs_da}
\addplot[thick, orange,dashed]  table [x=X, y=Y2, col sep=comma] {images/car.csv};\label{line:abs_z3}
\addplot[thick, green,dotted]  table [x=X, y=Y3, col sep=comma] {images/car.csv};\label{line:only_da}
\addplot[thick, purple,dashdotted]  table [x=X, y=Y4, col sep=comma] {images/car.csv};\label{line:only_z3}

\end{axis}
\end{tikzpicture}
		}
		\caption{Mountain Car} 
		\label{fig:m_car_graph}
	\end{subfigure}
	\hfill
	\begin{subfigure}[t]{0.24\linewidth}
		\centering
		\resizebox{\linewidth}{!}{ 
			\begin{tikzpicture}
\begin{axis}[
    xlabel={\tiny \bf Number of traces},
    ylabel={\tiny \bf Time (ms) -- log scale},
    width=1\textwidth,
    height=1\textwidth,
    legend style={at={(0.8,-0.01)},anchor=south,draw=none, font=\tiny},
    ymode=log, 
    xtick={20000,40000,60000,80000},
    xticklabel style={font=\tiny},
    yticklabel style={font=\tiny}
]

\addplot[thick,blue,solid]  table [x=X, y=Y1, col sep=comma] {images/datal_1.csv};
\addplot[thick, orange,dashed]  table [x=X, y=Y2, col sep=comma] {images/datal_1.csv};
\addplot[thick, green,dotted]  table [x=X, y=Y3, col sep=comma] {images/datal_1.csv};
\addplot[thick, purple,dashdotted]  table [x=X, y=Y4, col sep=comma] {images/datal_1.csv};

\end{axis}
\node at (1.1,2.6) [red,thick] {\tiny Time-Out};
\node at (2.3,2.5) [red,thick] {\tiny Time Out};
\end{tikzpicture}
		}
		\caption{Lunar Lander} 
		\label{fig:lun_graph_1}
	\end{subfigure}
	\hfill
	\begin{subfigure}[t]{0.24\linewidth}
		\centering
		\resizebox{\linewidth}{!}{ 

\begin{tikzpicture}
\begin{axis}[
    xlabel={\tiny \bf Number of traces},
    ylabel={\tiny \bf Time (ms) -- log scale},
    width=1\textwidth,
    height=1\textwidth,
    legend style={at={(0.8,-0.01)},anchor=south,draw=none, font=\tiny},
    ymode=log, 
    xtick={0,2000,4000,6000,8000},
    xticklabel style={font=\tiny},
    yticklabel style={font=\tiny}
]
\addplot[thick,blue,solid]  table [x=X, y=Y1, col sep=comma] {images/dataf16.csv};
\addplot[thick, orange,dashed]  table [x=X, y=Y2, col sep=comma]{images/dataf16.csv};
\addplot[thick, green,dotted]  table [x=X, y=Y3, col sep=comma] {images/dataf16.csv};
\addplot[thick, purple,dashdotted]  table [x=X, y=Y4, col sep=comma] {images/dataf16.csv};

\end{axis}
\end{tikzpicture}
		}
		\caption{F-16 Simulation Scenario 1 } 
		\label{fig:f16_graph}
	\end{subfigure}
	\hfill
	\begin{subfigure}[t]{0.24\linewidth}
		\centering
		\resizebox{\linewidth}{!}{ 

\begin{tikzpicture}
\begin{axis}[
    xlabel={\tiny \bf Number of traces},
    ylabel={\tiny \bf Time (ms) -- log scale},
    width=1\textwidth,
    height=1\textwidth,
    legend style={at={(0.8,-0.01)},anchor=south,draw=none, font=\tiny},
    ymode=log, 
    xtick={4000,8000,12000,16000,20000},
    xticklabel style={font=\tiny},
    yticklabel style={font=\tiny}
]
\addplot[thick,blue,solid]  table [x=X, y=Y1, col sep=comma] {images/dataf16_1.csv};
\addplot[thick, orange,dashed]  table [x=X, y=Y2, col sep=comma]{images/dataf16_1.csv};
\addplot[thick, green,dotted]  table [x=X, y=Y3, col sep=comma] {images/dataf16_1.csv};
\addplot[thick, purple,dashdotted]  table [x=X, y=Y4, col sep=comma] {images/dataf16_1.csv};

\end{axis}
\end{tikzpicture}
		}
		\caption{F-16 Simulation Scenario 2} 
		\label{fig:f16_graph_1}
	\end{subfigure}
	\caption{Comparison of four modes of our implementation for various case studies. The legend is as follows: 
    \protect\ref{line:abs_da} represents \textbf{Abs\_DA}, 
    \protect\ref{line:abs_z3} represents \textbf{Abs\_Z3}, 
    \protect\ref{line:only_da} represents \textbf{Only\_DA}, and  
    \protect\ref{line:only_z3} represents \textbf{Only\_Z3}.}
	\label{fig:comparison}
	\vspace{-3mm}
\end{figure*}

\subsection{Case Study 3: F-16 Autopilot MPC Controller~ \cite{ARCH18:Verification_Challenges_in_F_16}}
\label{sec:f16}

\revision{This benchmark models both the inner-loop and outer-loop controllers of the F-16 fighter jet.
We explore two scenarios. 
The first scenario involves reaching a specified altitude set point while maintaining a certain speed. 
The second scenario tests whether the automated collision avoidance system can recover the aircraft from a critical moment.
} 
\subsubsection{First Scenario}
\revision{In this scenario, the aircraft's goal is to reach a certain altitude while maintaining a specified speed within a timeline of $t$. 
There are 16 state variables (e.g., $\alt$, $\speed$, $\pit$, $\yaw$, $\roll$, $\pow$,  AoA (Angle of Attack) noted as $\aoa$, and etc).
Our exogenous variables are the initial settings for $\alt(0)$, $\aoa(0)$, $\speed(0)$, $\pit(0)$, and the power lag that the engine suffers ($\pow$). 
Our endogenous variables are $\alt(t)$, $\aoa(t)$, $\speed(t)$, $\pit(t)$, $\pow(t)$, and the actions of the autopilot system for $t>0$, which include changing the throttle $\throt(t)$ and adjusting the angle of the elevators $\elev(t)$ to control the pitch (nose up or down).
%
In this experiment, we investigate the actions ($\throt(t)$ and/or $\elev(t)$) that determine whether the plane succeeds or fails in reaching the desired checkpoint, achieving the desired speed, or violating aircraft limits such as upward acceleration, AoA, or minimum airspeed that could lead to stalling.
}
\subsubsection{Second Scenario} 
\revision{Here, we place the aircraft in a critical position near the ground to evaluate its collision avoidance system. 
This scenario involves using a larger set of variables, thereby increasing the dimensionality of our problem compared to the previous scenario. 
These critical moments involve high degrees of $\pit$, $\roll$, and $\yaw$, as well as low $\speed$ near the ground, which may lead to failures such as ground collision and violations of the aircraft's aerodynamic limits.
}
\revision{Our exogenous variables are the initial settings for $\alt(0)$, $\speed(0)$, $\pit(0)$, $\aoa(0)$, $\yaw(0)$, $\roll(0)$, and $\pow$, while the endogenous variables are $\alt(t)$, $\speed(t)$, $\pit(t)$, $\yaw(t)$, $\roll(t)$, for $t>0$ and the actions of the autopilot system. 
These actions include adjusting the degree of the rudder $\rud(t)$ to change the yaw of the plane, changing the degree of the aileron $\ail(t)$ to modify the roll of the plane, and controlling the throttle $\throt(t)$ and elevator $\elev(t)$.
As in the previous scenario, we are examining the autopilot decisions that influence whether the aircraft can successfully recover from a potential collision or avoid violating aerodynamic constrains. 
Additionally, we aim to identify the actual cause of the failures.}

\subsection{Performance Analysis}
\label{sec:analysis}

\revision{Figures~\ref{fig:m_car_graph},~\ref{fig:lun_graph_1},~\ref{fig:f16_graph}, and~\ref{fig:f16_graph_1} illustrate the results of our experiments for the Mountain Car, Lunar Lander, and both F-16 simulation scenario, respectively.}
Indeed \revision{all} graphs show \revision{a} similar profile in terms of \revision{the} behavior of the four modes of experiments 
mentioned in Section~\ref{sec:impl}.

As shown in the graphs, the abstraction algorithms (\textbf{Abs\_DA} and \textbf{Abs\_Z3}) 
demonstrate significantly better performance by orders of magnitude than the conventional solvers 
(\textbf{Only\_DA} and \textbf{Only\_Z3}), with the latter exhibiting exponential growth in runtime 
with an increasing number of traces. 
\revision{This demonstrates the effectiveness of our abstraction-refinement technique: it identifies the actual causes of failures while running much faster than techniques on concrete traces. 
As shown in Fig.~\ref{fig:lun_graph_1}, our technique processes up to 80,000 traces in under 250 seconds, whereas \textbf{Only\_Z3} times out with threshold 1200 seconds at 20,000 traces and \textbf{Only\_DA} at 55,000 traces. 
}

%
Notably, \textbf{Abs\_DA} outperforms \textbf{Abs\_Z3}, and \textbf{Only\_DA} shows better 
performance than \textbf{Only\_Z3}. 
This observation can be attributed to the fundamental differences between SMT solvers, which focus on logical consistency, \revision{and the searching methods developed in data analysis libraries, which are tailored for efficient searching in large datasets.}

\begin{table}[b]
	\vspace{-4mm}
	\caption{Experiment on 1000 traces }
\label{tab:ref}
	\centering
 \scalebox{.7}{
	
\renewcommand{\arraystretch}{1.2}

\begin{tabular}{|c||c|c|c||c|}

\hline
  {\bf Case Study} &    {\bf  Algorithm} &    $\boldsymbol{\alpha}$ & {\bf Refinement steps} & {\bf 
  Time (ms)} \\
\hline\hline
\multirow{6}{4em}{\bf Mountain Car} &\multirow{3}{4em}{Abs\_DA}  & 0.01 & 28 & 1024 \\
 & & 0.05 & 7 & 475 \\
 &  & 0.1 & 2 & 102 \\
\cline{2-5}
 & \multirow{3}{4em}{Abs\_Z3} & 0.01 & 22 & 9793 \\
 &  & 0.05 & 6 & 4757 \\
 &  & 0.1 & 2 & 1306 \\
\hline\hline
\multirow{6}{4em}{\bf Lunar Lander} & \multirow{3}{4em}{Abs\_DA} & 0.01 & 24 & 494 \\
 &  & 0.05 & 7 & 396 \\
 &  & 0.1 & 3 & 150 \\
\cline{2-5}
 & \multirow{3}{4em}{Abs\_Z3} & 0.01 & 19 & 2809 \\
 &  & 0.05 & 4 & 794 \\
 &  & 0.1 & 3 & 239 \\
\hline

\end{tabular}
 }
\end{table}
%
In Table~\ref{tab:ref}, we present a comparison between different valuations of the parameter 
$\alpha$, which represents the subset size of $\FunUndV$. 
We conducted an experiment to find an optimal value for $\alpha$. Our findings indicate that a very small $\alpha$ may require numerous refinements, as it needs to add more traces to identify the cause, which is inefficient. 
On the other hand, large values of $\alpha$ needs fewer refinements, but the under-approximation function has to process a larger amount of data, which increases the processing time. 
Therefore, there is a trade-off between the number of refinements and the total time spent on them.
We note that for row that have equal $\alpha$, we shuffle the trace set, which impact computing the 
under-approximation.

\subsection{Causality Analysis}
\label{sec:causalanalysis}

\revision{This section demonstrates an important aspect of this research in 
investigating the actual cause of safety failures in CPS to {\em explain} the underlying reason.
Our case studies involve simulations that specifically focus on the intersection of AI-enabled decision-making (Mountain Car and Lunar Lander), environmental dynamics feedback, and the correctness of a non-AI controller within an F-16 aircraft simulation.}

\begin{figure*}[t!]
    \centering
    \begin{minipage}{0.27\linewidth}
        \centering
        \resizebox{0.53\linewidth}{!}{\begin{tikzpicture}[node distance=2.3cm, thick]
    \centering
  \node[draw,rounded corners=5pt , minimum width=2.4cm, align=center] (square1) {\footnotesize$\pos_x(0)=0.00$\\\footnotesize$\pos_y(0)=1.41$ \\ \footnotesize$\action(0)=1$};
  \node[draw, rounded corners=5pt, minimum width=2.4cm, align=center, below of=square1] (square2) {\footnotesize$\pos_x(1)=-0.01$\\\footnotesize$\pos_y(1)=1.43$ \\ \footnotesize$\action(1)=3$};
  \node[draw,rounded corners=5pt , minimum width=2.4cm, align=center, below of=square2] (square3) {\footnotesize$\pos_x(i)=-0.32$\\\footnotesize$\pos_y(i)=0.15$ \\ \footnotesize \textcolor{black}{$\action(i)=1$}};
  \node[draw,rounded corners=5pt , minimum width=2.4cm, align=center, below of=square3] (square4) {\footnotesize$\pos_x(n)=-0.02$\\\footnotesize$\pos_y(n)=0.04$ \\ \footnotesize$\action(n)=0$};
  
  \draw[->] (square1) -- (square2);
  \draw[->, dotted] (square2) -- (square3);
  \draw[->,dotted] (square3) -- (square4);
  
  \node[above,yshift=0.5cm] at (square1.north) {Trace $\tau_0$};
  \node[yshift=-5mm] at (square4.south) {{$\neg \varphi_{\fail}$}};
  
  \node[draw,rounded corners=5pt , minimum width=2.4cm, align=center,right of=square1, node distance=3.2cm] (square1_right) {\footnotesize$\pos_x(0)=0.0$\\\footnotesize$\pos_y(0)=1.41$ \\ \footnotesize$\action(0)=3$};
  \node[draw, rounded corners=5pt, minimum width=2.4cm, align=center, below of=square1_right] (square2_right) {\footnotesize$\pos_x(1)=-0.01$\\\footnotesize$\pos_y(1)=1.42$ \\ \footnotesize$\action(1)=3$};
  \node[draw,rounded corners=5pt , color=red,minimum width=2.4cm, align=center, below of=square2_right] (square3_right) {\footnotesize$\pos_x(j)=-0.32$\\\footnotesize$\pos_y(j)=0.15$ \\ \footnotesize $\action(j)=2$};
  \node[draw,rounded corners=5pt , minimum width=2.4cm, align=center, below of=square3_right] (square4_right) {\footnotesize$\pos_x(m)=-0.36$\\\footnotesize$\pos_y(m)=0.05$ \\ \footnotesize$\action(m)=0$};
  
  \draw[->] (square1_right) -- (square2_right);
  \draw[->, dotted] (square2_right) -- (square3_right);
  \draw[->,dotted] (square3_right) -- (square4_right);
  
  \node[above,yshift=0.5cm] at (square1_right.north) {Trace $\tau_1$};
  \node[yshift=-5mm] at (square4_right.south) {\textcolor{red}{$\varphi_{\fail}$}};

\end{tikzpicture}}
        \caption{Simulated traces in Lunar Lander and causal effect of decision by the main engine.} 
        \label{fig:tarce_lun}
    \end{minipage}
    \hfill
    \begin{minipage}{0.34\linewidth}
            \centering
            \resizebox{\linewidth}{!}{\input{images/f16_analysis}}
            \caption{The F-16 scenario leads to failure due to a violation of the AoA limit.} 
            \label{fig:f16_analysis}
    \end{minipage}
    \hfill
    \begin{minipage}{0.34\linewidth}
            \centering
            \resizebox{\linewidth}{!}{\input{images/f16_analysis2}}
            \caption{The F-16 counterfactual scenario leads to success.} 
            \label{fig:f16_analysis_counter}
    \end{minipage}
    \vspace{-5mm}
\end{figure*}

\subsubsection{Mountain Car}

In Example \ref{ex:car_cause} (see Fig.~\ref{fig:tarce}), we prove that making a poor decision to 
accelerate to the right \revision{(i.e., $\action(0) = 1$)} leads to failing in reaching the mountain top 
(i.e., formula $\varphi_{\fail}$).
Instead, \revision{in the counterfactual scenario we observe that it is necessary to accelerate to the left to gain momentum in order to climb the mountain.}
This not only shows the earliest bad decision by the controller but also identifies the ``but-for'' 
scenario, meaning what would have happened if a different action was taken.
Additionally, counterfactual reasoning demonstrates how to fix the bad decision made by the neural network.

\subsubsection{Lunar Lander}

\revision{We observe that when there is a strong wind from left to right, some controllers tend to overuse the right engine, resulting in $\action=3$ during the initial steps. This causes the lander to drift to the left.
However, we observe that even in this situation where the lander is positioned to the left 
of the landing pad, the controller can use its left engine, $\action=1$, to move the lander to the right 
and land safely. However, some controllers use their main engine, $\action=2$, resulting in the 
lander not reaching the landing pad. This results in $\pos_x(t) < 0 - \epsilon$, constituting a failure.}

\revision{
To illustrate this further, Fig.~\ref{fig:tarce_lun} shows two traces starting from the same point but taking different actions in the first step.
Dotted transitions means the next state is not the immediate next time step. The final state of the 
traces $\tau_0$ and $\tau_1$ is the $n$th and $m$th state, respectively.
In trace $\tau_0$, $\action(0)=1$, while in trace $\tau_1$, $\action(0)=3$. 
However, in both traces, the controllers overuse the right engine in the initial steps (both controllers 
in $\tau_0$ and $\tau_1$ use the right engine $\action(1)=3$), causing the lander to drift far to the 
left, resulting in $\pos_x(i)=-0.32$ in $\tau_0$ and $\pos_x(j)=-0.32$ in $\tau_1$.
At this state, where in both scenarios the lander has the same position and setting, the controller in 
$\tau_1$ decides to use the main engine $\action(j)=2$, while the controller in $\tau_0$ opts to use 
the left engine $\action(i)=1$ to move the lander to the right. 
These decisions under similar conditions lead to the failure of $\tau_1$ (i.e., $\pos_x(m)=-0.36<0-\epsilon$) and the success of $\tau_0$ (i.e., $0-\epsilon<\pos_x(n)=-0.02<0+\epsilon$).
This finding indicates that the failure in $\tau_1$ using decision $\action(j)=2$, while the counterfactual scenario in $\tau_0$ succeeds with a different decision $\action(i)=1$, highlighting that $\action(j)=2$ in $\tau_1$ is the actual cause of the failure.}

\subsubsection{F-16 Autopilot Simulation}

\revision{
Here, we identify the cause of failures and analyze counterfactual scenarios (alternative actions) under the same conditions that could lead to success. 
%
%
\revision{When the aircraft needs to gain altitude at {\em low speed}, some traces show the controller lowering the nose to gain speed and avoid stalling before attempting to climb. 
This approach results in a loss of altitude and insufficient time to reach the desired altitude within the specified time frame, leading to failure. 
However, in counterfactual scenarios, the controller opts to gain speed by using more throttle and then gradually raises the nose using the elevators, eventually reaching the desired altitude. 
This demonstrates that the decision to lower the nose is the actual cause of the failure to reach the desired altitude within the specified time frame.}
 }

\revision{In another scenario, when transitioning from a lower to a higher altitude, some traces show controllers using excessive elevator and throttle, which places the aircraft in a danger zone and violates the angle of attach (AoA) limits, leading to catastrophic failure. 
However, in alternative counterfactual scenarios with the same starting conditions, the controller gradually uses the throttle and adjusts the elevator more cautiously. 
This approach allows the aircraft to reach the desired altitude without violating its aerodynamic limits.
}

\revision{
To illustrate the latter scenario in detail, Figs.~\ref{fig:f16_analysis} 
and~\ref{fig:f16_analysis_counter} show two flight real paths starting from the same altitude, 
$\alt(1)=1450$, and the same speed, $ \speed(1)=500$, with the goal of reaching an altitude of 
$\alt(n)=1800$. 
This process should occur within a specified time frame while not violating aircraft limits. 
In Fig. \ref{fig:f16_analysis}, the controller starts by using throttle $\throt(1)=0.64$ and setting the elevator to a negative position, $\elev(1)=-6$, to achieve a positive pitch angle. 
This decision continues in subsequent steps in a more extreme manner, with $\throt(2)=1.0$ (full throttle) and $\elev(2)=-25$, resulting in nearly a 45-degree pitch. 
Next, to counteract this situation, the controller attempts to use $\elev(3)=25$ and $\elev(4)=25$ to stabilize the aircraft's sharp nose-up attitude, leading to a negative AoA, $\aoa(5)=-17$.
Since the aircraft's maximum negative AoA limit is -15, $\aoa(5)=-17$ violates this limit, and the controller fails to achieve its objective. 
On the contrary, in the counterfactual scenario (see Fig.~\ref{fig:f16_analysis_counter}), the controller starting with less aggressive throttle and elevator adjustments, such as $\throt(1)=0$ and $\elev(1)=-6.8$, resulting in a slight pitch. 
This strategy continues similarly with $\throt(2)=0$ and $\elev(2)=-12$, avoiding harsh climbs to reach the destination. 
By examining this scenario, we find that in the first time step, Fig. \ref{fig:f16_analysis} makes the decisions $\throt(1)=0.64$ and $\elev(1)=-6.8$, while Fig.~\ref{fig:f16_analysis_counter} makes $\throt(1)=0.0$ and $\elev(1)=-6.8$ under the same conditions (same altitude, speed, and etc). 
This counterfactual example shows that an alternative decision by the controller leads to success, providing sufficient evidence that the initial decision is the actual cause of failure.
}

\vspace{-3mm}
\section{Related Work}
\label{sec:related}

There is a wealth of research on causality analysis in the context of embedded 
and component-based systems from different perspectives.
In~\cite{gs20,ga14,gss17,wg15,gmflx13,b24,gmr10}, a new structure of formal causal analysis is 
proposed that can serve as a substitute for the \HP causal model. 
This approach is distinct from our work, which utilizes a framework of causal analysis to identify the cause of a specific effect.
Recently, there has been great interest in using temporal logics to reason about causality and 
explaining bugs~\cite{cdffhhms22,fk17,cffhms22,bffs23}.
However, these lines of work either focus on only modeling aspects of causality or do not 
address the problem of scalability in automated reasoning about causality, which inherently 
involves a combinatorial blow up for counterfactual reasoning.
In the CPS domain, using causality to repair AI-enabled controllers has recently gained 
interest~\cite{lrcsl23}.
This work explored the construction of \HP models on AI-enabled controllers, the search for the 
cause of failure using a search algorithm, and the verification of these causes using \HP constraints.
In contrast, our work focuses on identifying the cause of failure efficiently in traces using \HP 
constraints and proposes an efficient method for doing so.
In~\cite{LeitnerFischer2012Towar-23317}, causal analysis is performed on system models and 
system execution traces. 
In contrast, our algorithm is designed to efficiently identify the cause of any potential failure. 
Additionally, our work is focused on systems such as CPS that interact with their environment.
Although the idea of abstracting causal models in terms of structural equations has been studied in 
the literature~\cite{rwbmjgs17,beh19,bh19}, these works do not attempt to establish a relation 
between {\em actual} causes in the abstract and concrete causal models.
In the studies~\cite{beh19,bh19,rwbmjgs17}, the concept of abstraction in causal models was 
introduced, along with the preliminaries required to construct an abstraction function that maps 
low-level variables to high-level variables. 
The work in \cite{rwbmjgs17} presents a more general form of abstraction, while \cite{beh19,bh19} focus on the concept of intervention in causal models and how to build an abstraction that preserves them. 
The distinction between our work and these studies lies in our objective; we are not aiming to construct causal models, but rather, we are utilizing abstraction to identify the cause of an effect in a more efficient manner.

In~\cite{cdffhhms22,ishar2009}, the concept of explaining counterexamples returned from the 
model checker is proposed, with one focusing on specifications in LTL format and the other in 
HyperLTL format. 
However, in our work, we aim to efficiently identify the cause of failure in an embedded system.

\section{Conclusion and Future Work}
\label{sec:concl}

We concentrated on designing an efficient technique to reason about actual causality.
We proposed an SMT-based 
formulation to determine whether for an input transition system or a 
set of traces and a state formula (the effect), there exists an actual cause.
 Since identifying an actual cause involves counterfactual reasoning and, hence, a combinatorial blow 
 up, we also introduced an efficient heuristic based on abstraction-refinement.
 We evaluated our techniques on \revision{three} case studies from 
 the CPS domain: AI-enabled controllers for a (1) Mountain Car, and (2) Lunar Lander~\cite{openaigym}, \revision{(3) and an MPC controller for an F-16 autopilot simulator~\cite{ARCH18:Verification_Challenges_in_F_16}.}
 
 One natural extension is to consider {\em probabilistic} actual causality, where 
 either occurrence of events in the system are associated with probabilities, or, data points follow 
 some distribution.
Another important direction is causal models where the system is partially observable.
This model would enable us to deal with unknowns while reasoning about the causal effect of events.
We are also planning to design a neurosymbolic methods by combining our SMT-based technique in this paper with machine learning approaches that predict causality.
 

\bibliographystyle{IEEETran}
\bibliography{bibliography}


\end{document}